\documentclass[11pt,preprint2]{aastex}

\usepackage{graphicx}
\usepackage{rotating}

\slugcomment{}

\shorttitle{The Spitzer AGN survey}
\shortauthors{Lacy et al.}

\begin{document}


\title{The {\em Spitzer} mid-infrared AGN survey. I - optical and near-infrared spectroscopy of candidate obscured 
and normal AGN selected in the mid-infrared.}


\author{
M.\ Lacy \altaffilmark{1},
S.E.\ Ridgway\altaffilmark{2}, 
E.L. Gates \altaffilmark{3},
D.M.\ Nielsen\altaffilmark{4}
A.O.\ Petric\altaffilmark{5},
A.\ Sajina\altaffilmark{6},  
T. Urrutia\altaffilmark{7}
S. Cox Drews \altaffilmark{8},
C. Harrison\altaffilmark{9},
N. Seymour\altaffilmark{10},
L.J.\ Storrie-Lombardi\altaffilmark{11}}
\altaffiltext{1}{National Radio Astronomy Observatory, 520 Edgemont
Road, Charlottesville, Virginia 22903}
\altaffiltext{2}{National Optical Astronomy Observatory, 950 North Cherry Avenue, Tucson, AZ 85719}
\altaffiltext{3}{UCO/Lick Observatory, P.O.\ Box 85, Mount Hamilton, 
CA 95140}
\altaffiltext{4}{Department of Astronomy, University of Wisconsin, 
475 N. Charter Street, Madison, WI 53706}
\altaffiltext{5}{Department of Astronomy, California Institute of Technology, Pasadena, CA91125}
\altaffiltext{6}{Department of Physics and Astronomy, Tuffs University, 212 College Avenue, Medford, MA 02155}
\altaffiltext{7}{Leibniz-Institut f\"{u}r Astrophysik Potsdam, An der Sternwarte 16, 14482, Potsdam, Germany}
\altaffiltext{8}{946 Mangrove Avenue \#102, Sunnyvale, CA 94086}
\altaffiltext{9}{Department of Astronomy, University of Michigan, Ann Arbor, MI 48109}
\altaffiltext{10}{CSIRO, P.O. Box 76, Epping, NSW 1710, Australia}
\altaffiltext{11}{Spitzer Science Center, California Institute of Technology,
Pasadena, CA91125}

\begin{abstract}
We present the results of a program of optical 
and near-infrared spectroscopic follow-up of candidate Active Galactic Nuclei (AGN) 
selected in the mid-infrared. This survey selects
both normal and obscured AGN closely matched in luminosity across a wide range, from Seyfert galaxies with bolometric
luminosities $L_{\rm bol} \sim 10^{10}L_{\odot}$, to highly
luminous quasars ($L_{\rm bol} \sim 10^{14}L_{\odot}$), and with redshifts from 0-4.3.
Samples of candidate AGN were 
selected through mid-infrared color cuts at several different 24$\mu$m 
flux density limits to ensure a range of luminosities at a given redshift.
The survey consists of 786 candidate AGN and quasars, 
of which 672 have spectroscopic 
redshifts and classifications. 
Of these, 137 (20\%) are type-1 AGN with blue continua, 
294 (44\%) 
are type-2 objects with extinctions $A_V \stackrel{>}{_{\sim}}5$
towards their AGN, 96 (14\%) are AGN with lower extinctions ($A_V \sim 1$) 
and 145 (22\%) have redshifts, but 
no clear signs of AGN activity in their 
spectra. 50\% of the survey objects have 
$L_{\rm bol} >10^{12}L_{\odot}$, in the quasar regime. We present composite
spectra for type-2 quasars and for objects with no signs of 
AGN activity in their spectra. 
We also discuss the mid-infrared -- emission-line
luminosity correlation and present the results of cross-correlations with 
serendipitous X-ray and radio sources. The results show that: (1) obscured objects dominate the overall 
AGN population,
(2) there exist mid-infrared selected AGN candidates which lack AGN signatures in their
optical spectra, but have AGN-like X-ray or radio counterparts, 
and (3) X-ray and optical classifications of obscured and unobscured AGN 
often differ.
\end{abstract}



\section{Introduction}

The past decade has seen a dramatic improvement in our ability to
find AGN that would be missing, or strongly selected against, in 
samples based on selection in the optical. 
There are several reasons why finding these objects is of interest. First,
the fraction of AGN obscured by dust represents a significant 
uncertainty in studies of AGN evolution. Second, the total number
of AGN (obscured plus unobscured) is needed to estimate the mean
efficiency of black hole accretion using the Soltan (1982) argument
(Martinez-Sansigre \& Taylor 2009). Third, 
some obscured AGN may represent an early phase in AGN activity,
as predicted by several models, and finding them would confirm the
importance of merger-driven evolution of massive galaxies, and give
strong clues about the nature of AGN feedback. Fourth, 
a sample of luminous AGN without a bright point source nucleus 
makes host galaxy studies of large numbers of these objects feasible.

Until 
the Sloan Digital Sky Survey (SDSS) and {\em Spitzer}, 
only a handful of high luminosity obscured AGN that were not either radio-loud 
(i.e.\ radio galaxies), or low luminosity (i.e.\ Seyfert-2 galaxies, 
rather than objects of quasar-like luminosity) 
were known. Norman et al.\ (2002) found an example radio-quiet type-2
quasar in the 
Chandra Deep Field South (CDFS), but the small areas of X-ray surveys mean
that relatively few high luminosity objects have been discovered in them. 
Radio-loud examples of this dusty type-1 quasar population
had been identified by Webster et al.\ (1998), but the degree to which 
synchrotron emission from jets associated with the radio source 
dominated the optical emission was unclear (Whiting et al.\ 2001).
Using the 2-Micron All-Sky Survey (2MASS) and near-infrared color 
selection, Cutri et al.\ (2001) were able to identify over 200 $z\sim 0.2$
radio-quiet red AGN, most likely reddened by dust (a 
similar technique, but using {\em Spitzer} 8$\mu$m photometry
as the long wavelength data
point, was used by Brown et al.\ 2006). Gregg et al.\ (2002) used a
combination of 2MASS and radio emission in the Faint Images of the Radio Sky 
at Twenty-cm (FIRST) survey 
to improve the reliability of selection of red quasars
by excluding non-AGN.  
These techniques typically found less heavily
obscured AGN ($A_V\sim 1-3$), predominantely broad-lined objects. 
This population of lightly dust-reddened AGN was found to contribute 
significantly ($\sim 20$\%) to the overall AGN population
at quasar-like luminosities, and has
properties consistent with an early phase in 
quasar evolution (Lacy et al.\ 2002; Glikman
et al.\ 2004, 2007; Urrutia et al.\ 2009, 2012). 
All of these early techniques, however, were still biased against true 
``type-2'' AGN with rest-frame $A_V\stackrel{>}{_{\sim}} 10$.

Narrow-line
selection of type-2 AGN from the Sloan Digital Sky Survey (SDSS) 
(Zakamska et al.\ 2003; Reyes et al.\ 2008) 
increased sample sizes of obscured, high luminosity AGN to several 
hundred at $z<0.8$, where the [OIII]5007 line is visible in the optical.
Mid-infrared selection of obscured AGN using {\em Spitzer} 
colors was then developed by
Lacy et al.\ (2004), Sajina, Lacy \& Scott (2005) and Stern et al.\ (2005).
In parallel, several other groups were working on joint radio/mid-infrared
selection (Martinez-Sansigre et al.\ 2005, Donley et al.\ 2005 and
Park et al.\ 2008), which gave more reliable, but less complete samples, 
missing radio-quiet objects. Mid-infrared techniques have 
found a significant population of both reddened
type-1 and luminous type-2 AGN at $z>1$ (e.g.\ Lacy et al.\ 2007a; 
hereafter L07), including 
the discovery of heavily absorbed objects invisible in all but the
deepest X-ray surveys.
Mid-infrared selected samples are not immune to 
problems; in particular they do not discriminate in principle
between dust heated
by an extremely dense, hot starburst, 
and an AGN power source
(so reliable samples require 
spectroscopic follow-up, as detailed in this paper), 
and the technique is not effective
at finding AGN which lack a hot dust component, such as low accretion rate
radio galaxies (e.g. Ogle et al.\ 2006). However, these samples do seem to
be complete for rapidly accreting, luminous AGN.

Using these early 
mid-infrared samples containing $\sim 10-100$ objects, 
Mart\'{i}nez-Sansigre et al.\ (2005) 
and L07 were able to show that dust-obscured 
AGN are at least as common as ``normal'' blue AGN, even at high 
luminosities. These conclusions
were also reached by Reyes et al.\ (2008) using SDSS optical selection
of type-2s, with a much larger sample (887 objects) at $z<0.8$.  
These samples were, however, too restricted
to examine trends of obscured AGN fraction 
as a function of both redshift and luminosity. Furthermore, the 
high redshift objects in the infrared-selected samples are, in general, 
significantly more luminous than those in X-ray samples of high redshift 
AGN, which are selected from relatively small fields
(tens of square arcminutes, compared to tens of square degrees). In these
less luminous AGN, {\em Hubble Space Telescope} 
images tend to reveal little disturbance in the
host, even at high redshifts (Kocevski et al.\ 2012), 
in contrast, the majority of dust
reddened quasars tend to have host galaxies showing strong signs of
disturbance (Lacy et al.\ 2007b; Urrutia et al.\ 2008). There thus 
may be a difference in the physical causes of AGN activity in 
Seyfert-like objects, powered by secular processes or minor 
interaction/merger activity in individual
galaxies, and quasars, powered by major galaxy-galaxy mergers 
(e.g.\ Treister et al.\ 2012).

We thus undertook a project to use the wide-area 
{\em Spitzer} Wide-area Infrared Extragalactic survey (SWIRE; Lonsdale
et al.\ 2003), the {\em Spitzer} Extragalactic First Look Survey (XFLS; 
Lacy et al.\ 2005; Fadda et al.\ 2006) and the {\em Spitzer} Cosmic Evolution
Survey (SCOSMOS; Sanders et al.\ 2007) 
surveys to search for luminous AGN in
an area of sky totaling 54 deg$^{2}$, large enough to find examples
of highly-luminous quasars at high redshifts. 
Furthermore, we nested our survey 
in terms of flux density limits at 24$\mu$m to ensure a good 
dynamic range in luminosity at a given redshift without the need 
to take many thousands of spectra. In 
this paper we present optical and near-infrared
spectroscopy of  our new survey. 
Our paper is structured as follows: 
Section 2 describes the selection of objects in the survey, 
how the individual sample
flux limits were chosen to ensure a wide range in AGN luminosity at a given 
redshift, and what selection effects remain. 
Section 3 describes the spectroscopic observations, 
Section 4 describes the classification of the objects, 
Section \ref{sec:xray}, the
X-ray detections in the survey, Section \ref{sec:radio}, the
radio detections, Section \ref{sec:type3}, the objects lacking 
AGN signatures in the optical, Section \ref{sec:oiii}, the
mid-infrared -- emission line luminosity correlation and Section 
\ref{sec:comp} the
composite spectra. Future 
papers (Ridgway et al.\ 2013; Lacy et al.\ 2013) will discuss 
the evolution and luminosity dependence of the demographics
of the obscured AGN population, and the spectral energy distributions
(SEDs) of the AGN. 

For the purposes of this paper, we define 
a quasar as an AGN having a total accretion luminosity of $>10^{12}L_{\odot}
(3.8\times 10^{45}{\rm ergs^{-1}cm^{-2}})$,  
which, assuming a bolometric correction to 15$\mu$m of nine from Richards et al.\ (2006)
for both obscured and unobscured objects, 
corresponds to ${\rm log_{10}}(L_{15\mu m} [{\rm ergs^{-1}Hz^{-1}]})>31.3$, or
${\rm log_{10}}(\nu L_{15\mu m} [{\rm ergs^{-1}}])>44.6$. (Note that this 
also corresponds approximately to $M_B {\rm (Vega)} < -23.0$ for an unreddened AGN.)
We assume a cosmology with $H_0=70 {\rm kms^{-1}Mpc^{-1}}$, $\Omega_M=0.3$
and $\Omega_{\Lambda}=0.7$.

\section{Survey selection}

\subsection{Infrared color selection criteria}

The candidate AGN were selected from the XFLS, SWIRE and SCOSMOS 
fields. Several samples were selected
limited at different 24$\mu$m flux density ranges to ensure a 
good spread of mid-infrared luminosities at a given redshift to aid with
disentangling correlations due to redshift from those due to luminosity.
Color selection based on the [3.6],[4.5],
[5.8] and [8.0] flux densities from the Infrared Array Camera (IRAC) was
then applied to each sample (Figure 1) 
in order to filter out low-$z$ starbursts 
and quiescent galaxies as detailed in L07 (except
with a slightly
expanded region in color space allowed for the fainter 24$\mu$m
samples, as discussed below). 

The L07 color ``wedge'' selection is as follows:
\begin{eqnarray}
\;\;\; {\rm log_{10}}(S_{8.0}/S_{4.5})& \leq & 0.8{\rm log_{10}}(S_{5.8}/S_{3.6})+0.5\nonumber\\
\&\; {\rm log_{10}}(S_{5.8}/S_{3.6})& > & -0.1 \nonumber\\
\&\; {\rm log_{10}}(S_{8.0}/S_{4.5}) & > & -0.2\nonumber\\
\end{eqnarray}
the expanded selection is:
\begin{eqnarray}
\;\;\; {\rm log_{10}}(S_{8.0}/S_{4.5})& \leq & 0.8{\rm log_{10}}(S_{5.8}/S_{3.6})+0.5\nonumber\\
\&\; {\rm log_{10}}(S_{5.8}/S_{3.6}) & > & -0.3 \nonumber \\
\&\; {\rm log_{10}}(S_{8.0}/S_{4.5}) & > & -0.3\nonumber\\
\end{eqnarray}

The samples are detailed in Table \ref{tab:samples}. 
Bright samples (those with flux 
density limits at 24$\mu$m ranging from $S_{24}>4$mJy to $>10$mJy) 
were selected across all 54 deg$^2$ of the
SWIRE fields, XFLS and SCOSMOS fields
in order to maximize the volume probed for 
very luminous objects. Faint samples, with limits ranging from 
$S_{24}>0.6$ to $>1.2$mJy,
were selected in small areas in individual SWIRE fields 
(totalling 2.2 deg$^{2}$) to provide 
a range in luminosity at every redshift. In addition, we selected
a sample with a very narrow range in $S_{24}$ ($1.0\leq S_{24}< 1.1$mJy)
split between three separate fields, 
which we followed up with Gemini Multi-Object Spectrograph 
on Gemini South (GMOS-S)
to improve our overall spectroscopic completeness at faint $S_{24}$
(the ``GMOSS'' sample in Table \ref{tab:samples}). Altogether, our
original selection included 963 candidate AGN (listed in Table 2), this was reduced
by constraints on fiber placement to 786 objects for which spectroscopy
was attempted, or had spectroscopic redshifts and type information in the literature (Table 3).

\begin{figure}
\plotone{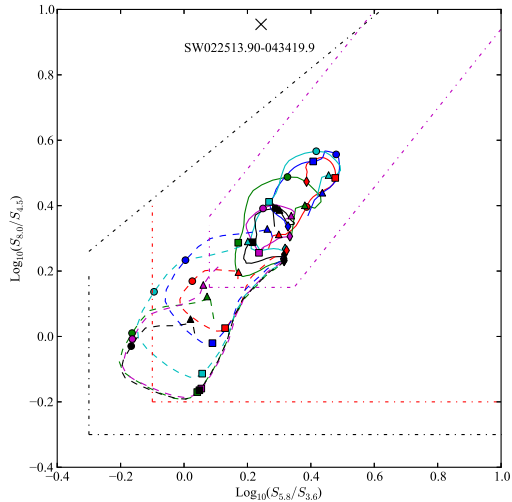}
\caption{Trajectories of six $z\sim 0.6$ type-2 quasars 
(from Lacy et al.\ 2007b, selected to have $S_{\rm [8.0]}>1$mJy) in 
color-color space. Solid lines represent
the unmodified fits to the SEDs, dashed lines the same objects, but
with the AGN mid-infrared component reduced by a factor of ten to 
represent objects close to the limit of our faintest samples. Symbols
represent redshifts of 0.3 (triangles), 1.0 (circles), 2.0 (squares) and
5.0 (diamonds). The red dot-dash line represents the L07 critera, 
the black dot-dash line the expanded wedge used for the faint samples, and
the magenta dot-dash line the AGN selection criteria of Donley et al.\ 
(2012). The
black cross near the top of the plot marks the position of SW022513.90-043419.9, 
a $z=3.43$ type-2 quasar discovered by Polletta et al.\ (2008) (see Section 
6 for discussion).}\label{fig:models}

\end{figure}

The selection region was changed from the L07 wedge (Eqn.\ 1) 
to Eqn.\ (2) for the fainter
samples to improve our completeness in low luminosity AGN, which 
tend to have bluer colors as starlight dominates the spectral energy 
distribution (SED) at rest 
wavelengths $\sim 2-4\mu$m rather than AGN dust emission. The expanded
criteria were determined by taking the SEDs of several well-studied type-2
quasars from Lacy et al.\ (2007b) and changing the redshifts and relative
contributions of the AGN components. This resulted in the trajectories 
in color space 
shown in Figure \ref{fig:models}, and led us to the conclusion that, to increase completeness,
we needed to expand the wedge.
Expanding the wedge in this way inevitably results in more contamination, but
the spectroscopy is able to remove the objects more likely to be starbursts 
from the final AGN sample.

\begin{figure*}
\begin{picture}(400,150)
\put(0,0){\includegraphics[scale=0.45]{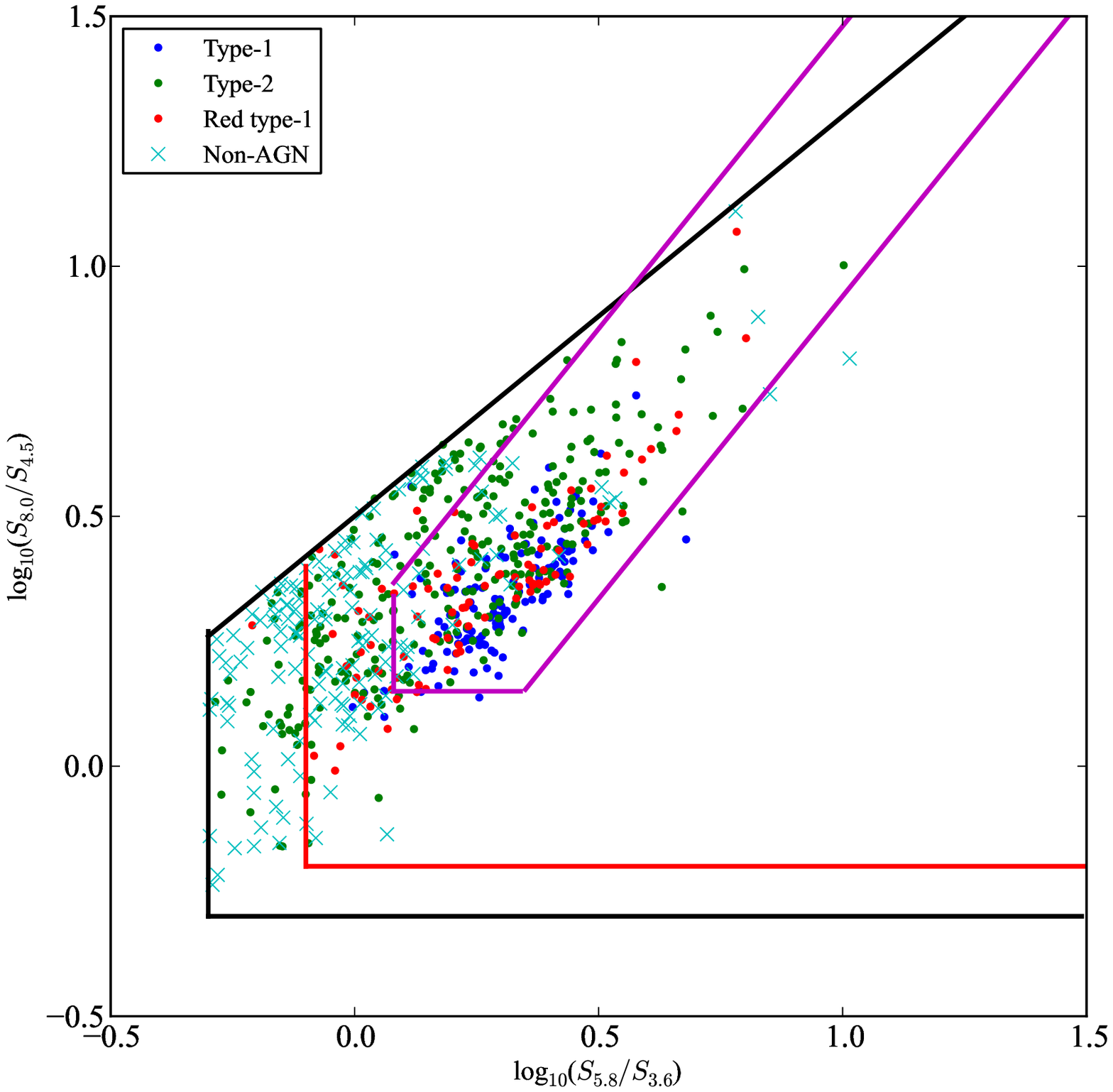}}
\put(230,0){\includegraphics[scale=0.45]{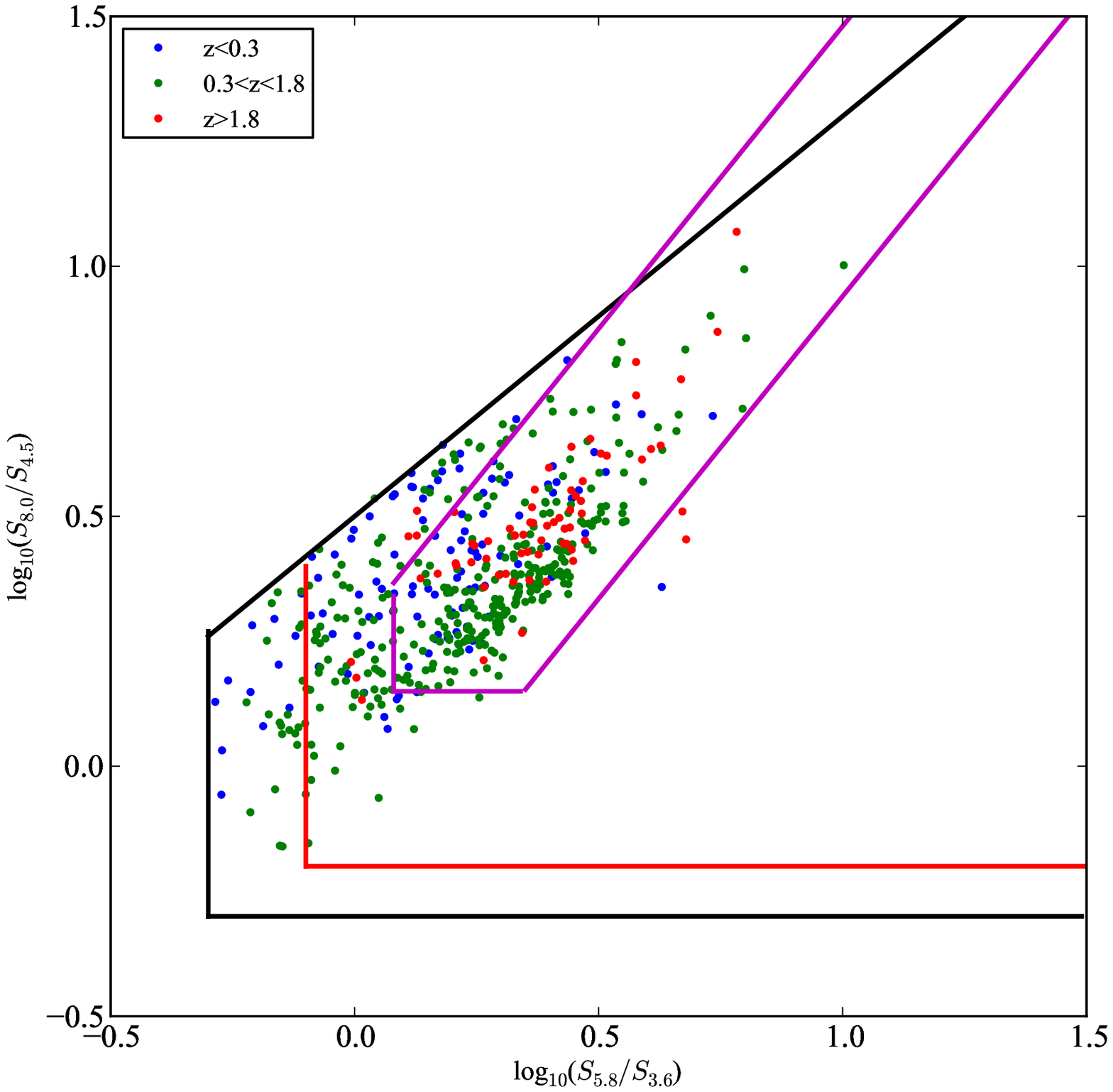}}
\end{picture}
\caption{Objects, colored according to (a) type (for all objects), 
and (b) redshift 
(for confirmed AGN only). 
The selection for the ``expanded'' wedge, used for the faint samples, is shown as the black solid line. The selection of L07, used for the brighter samples, is shown as the red solid  line. The selection critera of Donley et al.\ (2012)
is shown as the magenta solid line. Note that the Donley et al.\ 
criteria also require a monotonic increase in flux from [3.6] through [8.0].}\label{fig:wedge} 
\end{figure*}

Figure \ref{fig:wedge} shows the actual objects, color coded according to type
(see Section 4.2), in the 
selection region. In general, the AGN of all types lie close to the locus 
formed by the type-1s, however, many of the type-2s scatter above this line,
presumably due to the presence of polycyclic aromatic hydrocarbon (PAH) 
emission in the mid-IR spectra. There is also
a trend for high redshift objects to have redder colors in general.
Note that the expanded wedge selection did indeed make us more complete 
to low luminosity type-2s, though the overall reliability was relatively 
low within the expansion region - of 94 extragalactic objects which
lie only in the expanded region, 
51 (54\%) have non-AGN optical spectra, nine (10\%) have featureless spectra 
and 34 (36\%) have AGN spectra.

Various criteria for the mid-infrared color selection of AGN 
are compared and discussed in detail in Donley et al.\ (2012). They
show that it is possible to come up with better optimized selection for AGN
using IRAC colors, 
especially with regards to increased reliability of AGN selection. As 
Figure \ref{fig:wedge} shows, though, 
the strictness of the Donley et al.\ (2012) criteria does 
result in a significant
fraction of AGN being missed, particularly type-2s.

\subsection{Biases and incompletenesses}

We know of several selection effects that will remove objects from these
samples. In general, mid-infrared selection does 
not work well for objects with low accretion rates ($\stackrel{<}{_{\sim}}0.01$
Eddington, e.g.\ low luminosity radio galaxies [Ogle et al.\ 2006] and 
LINERS [Sturm et al.\ 2006]), 
whose mid-infrared luminosities are low compared to their bolometric 
output, perhaps because they lack the
torus structure of objects that accrete at a higher rate, and 
which is thought to be responsible for the mid-infrared emission 
from AGN. 

A second source of incompleteness was discussed in L07. It occurs 
at $z<0.3$, when the 7.7$\mu$m PAH feature from star formation is 
present in the IRAC 8.0$\mu$m band, and can be
of high enough equivalent width that the $S_{8.0}/S_{4.5}$ ratio 
is too high for the object to be classed as an AGN. Objects for which this
occurs are typically low luminosity objects, with weak
mid-infrared continua, as, for a given amount of PAH luminosity, the
equivalent width of the feature will be higher if the continuum is weak, 
however, our overall completeness is significantly  compromised 
at those redshifts (by perhaps as much as 50\% based on L07). The effects
of this on our estimate of the AGN luminosity function will be discussed
in Ridgway et al.\ (2013).

A third
possible selection effect is related to the highest redshift objects only, 
where the colors become more difficult to predict, and 
the requirement for four-band IRAC detection means that some extreme 
objects may be missed. Although we have
succeeded in finding several high redshift objects, both of the 
$z\sim 3.5$ objects of Polletta et al.\ (2008) would have been missed
from the color selection due to their very steep mid-infrared spectra at 
$>5 \mu$m in the observed frame (indeed, one of them, 
SW0225550.32-042149.6, 
is not even detected at 5.8$\mu$m in SWIRE). Objects with 
extremely low ratios of stellar host luminosity to AGN hot dust luminosity are 
not represented in our models in Figure 1 (or in Sajina et al.\ 2005), but
seem to be present at high redshift. The redshift range 
$z\sim 3-4$ is likely to be particularly problematic, as the 
5.8$\mu$m/3.6$\mu$m (rest frame $\sim$1.3$\mu$m/0.8$\mu$m) color 
is dominated by stellar emission from the host, with
only a small contribution from the AGN power-law, and is thus
relatively blue, whereas
the 8.0$\mu$m/4.5$\mu$m (rest-frame 1.8$\mu$m/1.0$\mu$m) color can be very 
red if the AGN is powerful and dominates the 8.0$\mu$m flux, as the 4.5$\mu$m
emission will be dominated by stellar emission from the host. At $z<3$, the 
effect will be less pronounced. The 5.8$\mu$m/3.6$\mu$m will be redder
because more of the AGN emission is in the 5.8$\mu$m band, and the
object will move into the selection wedge (to the right in Figure 1). 
Similarly, at $z>4$, the 8.0$\mu$m emission 
from the AGN will be more diluted by the stellar population of the host, 
rendering the 8.0$\mu$m/4.5$\mu$m color bluer, moving the object down
into the selection wedge in Figure 1. 
A related source of bias against high redshift objects is that
high redshift objects with very faint host galaxies would have been missed
at the shorter wavelengths, where the host galaxy dominates the SED.

We can quantify some of these incompleteness effects with the aid of the new 
{\em Spitzer} Extragalactic Representative Volume Survey (SERVS; Mauduit 
et al.\ 2012), which provides deep IRAC 3.6 and 4.5$\mu$m data over 
a total of 18 deg$^{2}$ spread over several SWIRE fields. First, we examined how many
reddened, high redshift AGN we may be missing due to their falling outside
of our selection wedge (e.g.\ SW022513.90-043419.9 in Figure 1). In the 4 deg$^2$ Lockman
Hole SERVS field, we find 38 objects with $S_{24}> 1$mJy (comparable to our
``Deep'' sample flux limits) that have ${\rm log_{10}}(S_{5.8}/S_{3.6})>0.1$ 
and lie above the diagonal line in Figure \ref{fig:models} (and therefore
are candidates for very high redshift objects missed by our selection wedge).
Only 13 of these, however, have $r>20$, consistent with them being $z>0.3$
objects missed by our selection. The ``Deep'' fields in this paper contain
405 objects (with 90\% redshift completeness) within a comparable area 
(3.2deg$^2$). The addition of $\sim 10$ more high redshift objects would 
thus not severely affect the completeness of the sample, though losing these may 
lead to a bias against the highest redshift obscured quasars. Second, 
we examine how many objects may be missing due to not being detected
in the shortest wavelength 
IRAC bands given the SWIRE and XFLS flux density limits. By combining 
with SERVS 3.6 and 4.5$\mu$m data, we were able to verify that the faintest 
object at 3.6$\mu$m, with $S_{3.6}=8\mu$Jy, is above the flux density limit
of SWIRE ($S_{3.6}=7.3\mu$Jy), though below that of the XFLS 
($S_{3.6}=20\mu$Jy). Given the density of sources with $S_{3.6}<20\mu$Jy,
we estimate that four objects may be missing from our XFLSDeep sample.



\section{Observations}

\subsection{Optical spectroscopy}

A wide range of optical facilities and instruments were used for 
spectroscopic follow-up of our AGN candidates. 
Most of the bright samples were followed
up with 3-5m telescopes and longslit spectroscopy (Hale with COSMIC, 
SOAR with Goodman, Shane with Kast), whereas the fainter samples 
were followed up with multifiber and/or 6-8m class telescopes 
(Blanco with Hydra, MMT with Hectospec and Gemini-South with
GMOS [program GS-2008B-C4]). 
We also obtained spectra of some of the bright candidates with a successful 
poor weather (scheduling band 4) program at Gemini-South 
(program GS-2008B-Q86).
Some objects had spectra available in archives from the SDSS, 
2dF (Colless et al.\ 2001) or 6dF (Jones et al.\ 2009) 
surveys, and some have redshifts and
classifications in the literature, all found using the NASA Extragalactic 
Database (NED).
Table 2 (full table in electronic format) gives details of the 
spectroscopic observations or literature references as appropriate.
Note that we include the
previously--published spectra of L07 for completeness. (For reasons of 
space the optical spectra are not shown in this paper, however, we intend to
make them available electronically, either by direct application to the 
author, or through a data service.) 

Data analysis followed standard procedures. Most of the data were analysed in IRAF, using the twodspec package to 
apply bias subtraction, flat fielding, wavelength calibration and extraction. Correction for atmospheric extinction 
was performed using mean extinction curves appropriate to the observatory. No correction has been made to the
spectra for Galactic extinction. 

The fiber data from the Hydra instrument was analysed in the hydra package of IRAF, 
with improved sky subtraction using our implemention of an algorithm used at the Anglo-Australian Telescope. This involved median filtering each fiber spectrum with a broad (201 pixel, 168\AA ) filter, 
subtracting the median filtered continuum, then measuring the flux in the prominent sky lines at 5577\AA, 5892\AA,
6300\AA, 6363\AA$\;$and 6832\AA. The fluxes of the same lines in the combined sky
spectum were then matched to these, and a mean scaling factor found, which was then applied to the sky spectrum
before it was subtracted from the object spectrum. 

The MMT Hectospec data were analysed using 
software based on the {\sc hsred} package of R.\ Cool,
with the same modification disussed above to improve the sky subtraction.  
In most cases our sky subtraction technique works well, particularly for the 
discrete sky lines. Remnant sky lines were masked out in the final spectra.
Fiber spectra of faint objects 
are, however, more vulnerable to 
artifacts due to poor sky subtraction than longslit spectra. 
In cases of low signal-to-noise, we therefore
attempted to obtain confirmation via near-infrared spectroscopy 
(Section 3.1), or
used SED fitting to check that the redshifts we obtained were consistent
with the photometry (see Lacy et al.\ 2013).

\begin{sidewaystable}
\caption{Samples used}
{\scriptsize
\begin{tabular}{lcclcclccc}
Name           & Field            & $S_{24}$ range  & Effective              &$N^{a}$ & selection   & follow-up                &completeness   &90\% flux&$N_{90}^{b}$\\  
               & center(s)           & (mJy)           & area (sr)              &    &             &                          &($q^{c}\leq 3$)(\%)& limit   &        \\
               &            &                       &                        &    &             &                          &               &         &        \\\hline
ELAIS-S1Deep$^{d}$    &003444-4328&$1.30\leq S_{24}< 5.8$ & 0.000103&27& Exp.\ wedge& CTIO/Hydra    & 15  &2.98 &4 \\
ELAIS-S1Bright$^{d}$ & 003830-4400&  $S_{24}\geq 5.8$     &  0.002071              & 37 & Exp.\ wedge & SOAR/Goodman; Gemini/GMOS& 100&6.1 & 36\\
XMM-LSSBright$^{d}$  &022120-0430&  $S_{24}\geq 6.6$           &  0.002772              & 34 & L07 wedge   & Hale/COSMIC              &100&6.6 & 34\\
XMM-LSSDeep$^{d}$ &021910-0500&  $1.19\leq S_{24}< 6.6$&  0.0000879& 52 & Exp.\ wedge & CTIO/Hydra               & 33&2.35& 17\\ 
CDFSBright$^{d}$    & 033200-2816&  $S_{24}\ge 6.0$      &  0.002376              & 40 &  L07 wedge  & SOAR/Goodman; Gemini/GMOS&100    &6.0 & 40\\
GMOS-S$^{d}$         & 003254-4253,  & $1.0\leq S_{24} < 1.10$& 0.000318 & 36 & Exp. wedge & Gemini-S/GMOS; CTIO/Hydra & 81 & N/A & N/A\\
                &003444-4328,  &                        &          &   &            &\\
                &021910-0500   &                        &          &   &            &\\
SCOSMOSBright$^{e}$   & 100000+0200& $S_{24}\geq  6.3$ & 0.0006092                        & 8 & L07 wedge   & SOAR                     & 100 &    6.3 &   8\\ 
LockmanBright$^{d}$ & 104500+5800&  $S_{24}\geq 8.0$     &  0.003351              & 28 & L07 wedge   & Hale/COSMIC              & 96&8.0 & 27\\ 
LockmanDeep$^{d}$   & 105248+5730&  $1.0\leq S_{24}<8.0$ &  0.0002175             & 133& Exp.\ wedge & MMT/Hectospec            & 90&1.03 &116\\
ELAIS-N1Deep$^{d}$  & 161024+5436&  $1.0\geq S_{24}< 8.0$     &  0.0002132             & 132& Exp.\ wedge & MMT/Hectospec            & 86&1.306& 76\\
ELAIS-N1Bright$^{d}$& 161100+5500&  $S_{24}\geq 8.0$     &  0.002815              & 19 & L07 wedge   & Lick/Kast & 88 & 9.6 & 13\\
ELAIS-N2Bright$^{d}$& 163648+4102&   $S_{24}\geq 8.0$    &  0.001277              & 15 &  L07 wedge  & Lick/Kast & 100& 8.0 & 15\\
XFLSBright$^{f}$    & 171800+5930&  $S_{24}\ge 4.6$      &  0.001157              & 46 & L07 wedge   & mix, see L07             &100&4.6 & 45\\
XFLSDeep$^{f}$       & 171517+5956&$0.61\leq S_{24}< 4.6$ &  0.0002735             & 178& Exp.\ wedge & MMT/Hectospec            & 91&0.61&175\\
\end{tabular}

\noindent
Notes:-

\noindent
$^a$ number of objects with either spectroscopy attempted, or literature redshifts.

\noindent
$^b$ number of objects in the 90\% complete samples.

\noindent
$^c$ redshift quality (see Section 4.1)

\noindent
$^{d}$ {\em Spitzer} photometry from SWIRE (Lonsdale et al.\ 2003)

\noindent
$^{e}$ {\em Spitzer} photometry from SCOSMOS (Sanders et al.\ 2007)

\noindent
$^{f}$ {\em Spitzer} photometry from the Extragalactic First Look Survey (Lacy et al.\ 2005; Fadda et al.\ 2006)

}\label{tab:samples}

\end{sidewaystable}

%
%
%
%
%
%
%
%
%
%


\begin{sidewaystable}
{\tiny
\caption{Mid-IR-selected AGN candidates and follow-up spectroscopy log}
\begin{tabular}{llcclccccccccccccccc}
Object Name & Sample &  Telescope/               & Observation &Exposure & MagSys$^{a}$&$m_1$&$m_2$&$m_3$&$m_4$&$m_5$&oflg$^{b}$ &oref$^{c}$ &$Z$&$Y$&$J$&$H$&$K$&nflg$^{d}$&nref$^{e}$ \\
            &           &  Instrument               & Date (UT)   & Time &             &     &     &     &     &     &           &           &   &   &   &   &   &            &             \\ 
            &           &  (or literature reference)  &             &(s) &             &     &     &     &     &     &           &           &   &   &   &   &   &            &             \\\hline
SW002644.79-430958.1  &     ELAIS-S1Bright& SOAR/Goodman   &               2008-08-01UT    &          900  &Vega:UBVRI&&19.47&&     &&29&APM &&&&&&64&  \\  
SW002802.45-424913.5  &     ELAIS-S1Bright& 6dF            &                               &               &Vega:UBVRI&&     &&17.6 &&23&6dF &&&&&&64&  \\
SW002802.79-425957.0  &     ELAIS-S1Bright& SOAR/Goodman   &               2008-08-03UT    &  3$\times$900 &Vega:UBVRI&&20.28&&     &&29&APM &&&&&&64&   \\
SW002927.72-431614.4  &     ELAIS-S1Bright& Gemini-S/GMOS  &               2008-10-10UT    &  3$\times$900 &          &&     &&     &&64&    &&&&&&64&  \\
SW002933.86-435240.4  &     ELAIS-S1Bright& LF04           &                               &               &Vega:UBVRI&&     &&17.7 &&23&LF04&&&&&&64&   \\
SW002959.22-434835.1  &     ELAIS-S1Bright& LF04           &                               &               &Vega:UBVRI&&     &&17.42&&23&LF04&&&&&&64&  \\
SW003114.45-424227.7  &     ELAIS-S1Bright& LF04           &                               &               &Vega:UBVRI&&     &&20.1 &&23&LF04&&&&&&64&  \\
SW003224.37-424211.0  &     GMOS-S        & Gemini/GMOS-S  &               2008-10-02UT    &   2$\times$600&          &&     &&     &&64&    &&&&&&64&  \\\hline
\end{tabular}

\noindent
Notes:-Table 2 is published in its entirety in the electrononic edition of Astrophysical Journal Supplement, a portion is shown here for guidance regarding its form and content. 
Note that we also include sources that satisified our selection criteria, but which
lack spectroscopic observations for completeness. Literature references: 2dF - Colless et al.\ 2001; 6dF - Jones et al.\ 2009;   FBQS - White et al.\ 2000; LF04 - La Franca et al.\ 2004; Mao12 - Mao et al. (2012); P06 - Papovich et al.\ (2006); 
RC3 - de Vaucouleurs, G.\ et al.\ 1991; S01 - Serjeant et al.\ 2001; S-A94 - di Serego-Alighieri et al.\ 1994; 
S06 - Simpson et al.\ 2006; S12 - Simpson et al (2012); Sh02 - Sharp et al.\ 2002.

\noindent
$^{a}$ Optical magnitude system and filters: AB or Vega denotes the system, followed by a colon, 
then the filter set used in the next 
five columns ($m_1-m_5$) (i.e.\ Vega:UBVRI or AB:ugriz). 
(Note that all the near-IR magnitudes, $Z,Y,J,H,K$, are in the Vega
system.)

\noindent
$^{b}$ Bit flag indicating which optical magnitudes are limits or missing, values are added to the 
bit flag as follows: 1-$m_1$ is a limit/missing, 2-$m_2$ is a 
limit/missing, 4-$m_3$ is a limit/missing, 8-$m_4$ is a limit/missing, 16-$m_5$ is a limit/missing. For
example, a bit flag of 17 (1$+$16) indicates that $m_1$ and $m_5$ are missing/limits, and the remainder of 
$m_2,m_3$ and $m_4$ are detections. Where limits are known, they are given, otherwise the 
magnitude column is left blank. An oflg of 64 indicates that no optical detection has been reported for this 
object.

\noindent
$^{c}$ Source of the optical data. APM - Maddox et al.\ (1990); 
CASU - Gonz\'{a}les-Solares et al.\ (2011); CFHTLS - Gwyn (2012); 
COSMOS - Sanders et al. (2007); ESIS - Berta et al.\ (2006); LCRS - Schechtman et al.\ 1996; MS4 - Burgers \& 
Hunstead (2006); 
SDSS - Abazajian et al.\ 2007; SWIRE - Lonsdale et al. (2003; http://swire.ipac.caltech.edu); USNO-A2 - http://ftp.nofs.navy.mil/projects/pmm/catalogs.html). Others as for the
spectroscopic references above. ``This paper'' refers to estimates from spectroscopic 
acquisition images, or by-eye estimates
from available imaging, with an uncertainty $\approx 0.5$ magnitudes. The other magnitudes are accurate to 
$\approx 0.1$ magnitudes or better, see the respective papers for details. CASU and SWIRE magnitudes are
aperture magnitudes in 2.4 or 3.1 arcsecond diameters, respectively. The remainder are estimated
total magnitudes (``MAG\_AUTO'' in Sextractor, Bertin \& Arnouts [1996]).

\noindent
$^{d}$ Bit flag indicating which near-infrared flags are limits or missing, values are added to the 
bit flag as follows:: 1-$Z$ is a limit/missing, 2-$Y$ is a 
limit/missing, 4-$J$ is a limit/missing, 8-$H$ is a limit/missing, 16-$K$ is a limit/missing. The resulting 
nflg is then calculated by summing these values. Where limits are known, they are given, otherwise the 
magnitude column is left blank. An nflg of 64 indicates that no near-infrared detection has been reported for 
this object.

\noindent
$^{e}$ Source of the near-infrared data. 2MASS - Strutskie et al. (2006); 
DXS, UDS - the Deep Extragalactic Survey and Ultra-Deep Survey
of the United Kingdom Infrared Deep Sky Survey (UKIDSS) (Lawrence et al.\ 2007). VIDEO - Jarvis et al.\ (2013). 
For DXS and UDS the 2 arcsecond diameter apertures are 
quoted, for VIDEO the Petrosian magnitudes. ``This paper'' refers to estimates from spectroscopic
acquisition images, with an uncertainty $\approx 0.5$ magnitudes. The other magnitudes are accurate to 
$\approx 0.1$ magnitudes or better, see the respective papers for details. Note that all near-infrared
magnitudes are on the Vega system.

}
\label{tab:obslog}
\end{sidewaystable}

\begin{sidewaystable}
{\scriptsize
\caption{Properties of the AGN in the spectroscopic survey.}
\begin{tabular}{lccccccccccc}
Object Name & Sample &In &  Redshift & Redshift & Type$^b$&Type   &$S_{\rm 3.6}$&$S_{\rm 4.5}$&$S_{\rm 5.8}$&$S_{\rm 8.0}$ &$S_{\rm 24}$ \\
            &           &statistical&           & quality$^a$&     &quality$^c$&($\mu$Jy)    &($\mu$Jy)    &($\mu$Jy)    &($\mu$Jy)    &($\mu$Jy)   \\ 
            &           &sample?&           &          &     &       &             &             &             &             &            \\\hline			     
SW002644.79-430958.1   &    ELAIS-S1Bright& Y &  0.241 &1 &  3 &  2 &  393.79&   480.71&    605.92&   1890.67 &  9629.14\\ 
SW002802.45-424913.5   &    ELAIS-S1Bright& Y &  0.127 &1 &  1 &  1 & 1260.01&  1529.63&   2171.13&   3210.09 & 13819.66 \\
SW002802.79-425957.0   &    ELAIS-S1Bright& Y &  1.731 &2 &  4 &  2 &  464.41&   842.38&   1655.11&   3256.47 & 11582.89 \\
SW002927.72-431614.4   &    ELAIS-S1Bright& Y &  0.593 &1 &  2 &  1 &  307.22&   409.86&    647.29&    940.93 &  5804.15\\ 
SW002933.86-435240.4   &    ELAIS-S1Bright& Y &  0.994 &1 &  1 &  1 &  842.44&  1183.64&   1701.33&   2374.71 &  7102.62 \\
SW002959.22-434835.1   &    ELAIS-S1Bright& Y &  2.039 &1 &  1 &  1 &  571.40&   832.07&   1501.68&   2611.84 &  6321.39\\ 
SW003114.45-424227.7   &    ELAIS-S1Bright& Y &  0.593 &1 &  1 &  1 & 1108.38&  1744.93&   2532.70&   4025.79 & 11810.21 \\
SW003119.19-424533.9   &    GMOS-S        & Y &  0.494 &1 &  3 &  1 &  184.00&   150.00&    139.00&    235.00 &  1013.00\\\hline 
\end{tabular}

Notes:-
\noindent
Redshift quality, type and type quality flags are discussed in detail in Sections 4.1 and 4.2, and are summarized 
here:

\noindent
$^a$ redshift quality: 1-secure redshift based on two or more high signal-to-noise features; 2-less secure 
redshift, based on multiple features, but with only one or fewer detected at high signal-to-noise; 3- uncertain
redshift, based on weak spectral features, or a single strong line; 4- featureless spectrum, no redshift estimate. 

\noindent
$^b$ type: 1- normal, unobscured type-1 AGN; 2- heavily obscured, type-2 AGN; 3- no indication of an AGN in the
optical spectra; 4- lightly obscured AGN, with broad-lines visible in the rest-frame optical, but a red continuum.
\noindent

$^c$ type quality: 1- secure classification (broad lines for type-1s, 
BPT diagram, [NeV] emission or high-ionization UV lines and a rest-frame optical spectrum for type-2s; 
clear $(g-i)^*$ color excess compared to normal quasars); 
2- less secure classification (e.g.\ only
partial information for BPT, low signal-to-noise high-ionization line detections); 3 or 4- uncertain classification
due to lack of strong spectral features and/or uncertain line identification.

\noindent
Table 3 is published in its entirety in the electrononic edition of Astrophysical Journal Supplement, a portion is shown here for guidance regarding its 
form and content.
}\label{tab:redshifts}
\end{sidewaystable}

In the case of the fiber spectrographs, $\approx$ 20\% of the candidates did not have spectra taken as 
we only used a single fiber configuration
for each field. In some cases fibers could not be placed close enough 
together to obtain spectra for all objects (the minimum fibre separation
was 20 arcseconds for Hectospec and 25 arcseconds for Hydra), in others objects 
were not observed in order to prevent fibers crossing. 
For these fields, the effective area was calculated as the area of the survey multiplied by the ratio of the number of objects for which spectra were attempted plus those
which had spectra in the literature (and were therefore excluded from the fiber assignments) to the 
total number of candidates in the area and flux range. Exclusions due to fiber crowding were 
randomly determined to avoid possible bias.

\subsection{Near-infrared spectrosopy}

\begin{figure}
\plotone{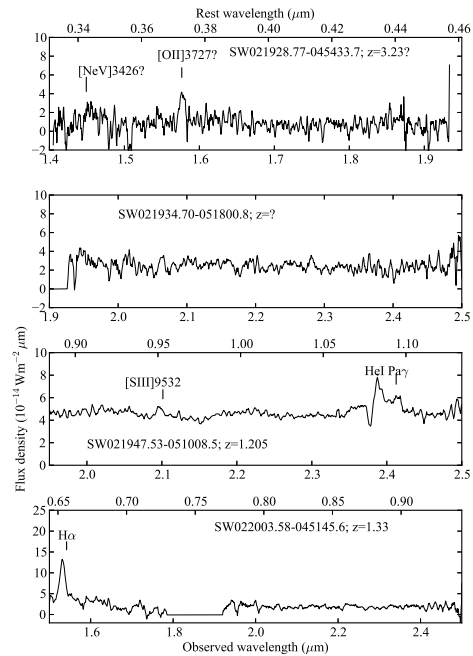}

\caption{Near-infrared spectra of high redshift objects in the
XMMDeep sample.}\label{fig:nir_xmm}
\end{figure}

\begin{figure}
\plotone{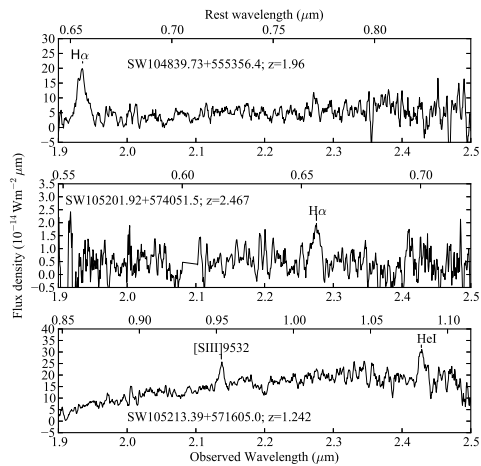}

\caption{Near-infrared spectra of high redshift objects in the
LockmanBright and LockmanDeep samples.}\label{fig:nir_lh}

\end{figure}

\begin{figure}
\plotone{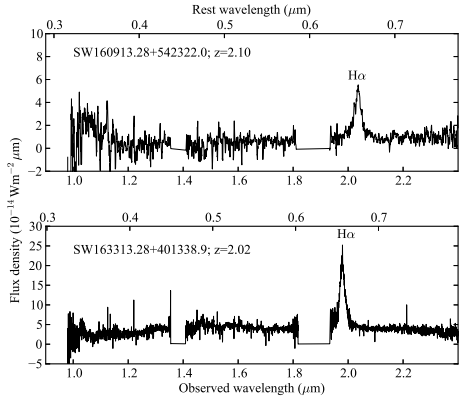}

\caption{Near-infrared spectra of high redshift objects in the
ELAIS-N1Bright and ELAIS-N2Bright samples.}\label{fig:nir_en}

\end{figure}

\begin{figure}
\plotone{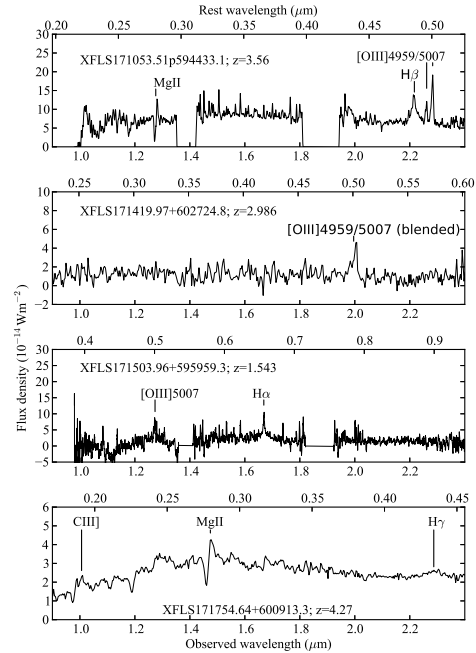}

\caption{Near-infrared spectra of high redshift objects in the
XFLSBright and XFLSFaint samples.}\label{fig:nir_xfls}

\end{figure}

For some high redshift candidates with ambiguous or low signal-to-noise 
optical spectra we were able to obtain 
near-infrared spectra with IRTF using Spex (Rayner et al.\ 2003) , 
Gemini with NIRI (program GN2009B-C-8) and Triplespec (Herter et al.\ 2008)
on Palomar. The near-infrared 
observations are listed in Table \ref{tab:nir}, and the 
corresponding
spectra are shown in Figures \ref{fig:nir_xmm}-\ref{fig:nir_xfls}.  

\begin{table*}
\caption{Near-infrared spectroscopy}\label{tab:nir}
{\scriptsize
\begin{tabular}{lcclrl}
Object &  Subsample & Telescope& Instrument/Band & Integration time & Observation date\\
       &            &          &                 &                  &                 \\\hline
SW 021928.77-045433.7  &XMM-LSSDeep   & Gemini-N  & NIRI/H    &37x120&2009-11-29\\
SW 021934.70-051800.8  &XMM-LSSDeep   & Gemini-N  & NIRI/K    &37x120&2009-11-28\\
SW 021947.53-051008.5  &XMM-LSSDeep   & Gemini-N  & NIRI/K    &13x120,18x120 & 2009-12-08,09\\
SW 022003.58-045145.6  &XMM-LSSDeep   & Gemini-N  & NIRI/J,K  &31x120,37x120 &2009-11-29,2009-11-30\\
SW 022003.58-045145.6  &XMM-LSSDeep   & Hale      & Triplespec&8x300&2008-07-27\\
SW 104839.73+555356.4  &LockmanBright& Gemini-N & NIRI/K    &26x120&2009-11-29\\
SW 105201.92+574051.5  &LockmanDeep & Gemini-N  & NIRI/K    &25x120&2009-11-28\\
SW 105213.39+571605.0  &LockmanDeep & Gemini-N  & NIRI/K    &19x120&2009-11-28\\
SW 160913.28+542322.0  &LockmanDeep & Hale      & Triplespec&20x300&2011-07-12\\
SW 163313.28+401338.9  &EN2Bright   & Hale      & Triplespec&12x300&2011-07-11\\
XFLS 171053.51+594433.1&XFLSFaint  & Hale      & Triplespec &24x300&2011-07-11\\
XFLS 171419.9+602724   &XFLSBright   & IRTF      & Spex     & 30x120& 2007-06-18\\
XFLS 171503.96+595959.3&XFLSFaint  & Hale      & Triplespec &24x300 &2011-07-12\\
XFLS 171702.54+600620.7&XFLSFaint  & Hale      & Triplespec &16x300 &2011-07-11\\
XFLS 171754.6+600913   &XFLSBright   & IRTF      & Spex     & 30x120&2007-06-19\\
\end{tabular}
}
\end{table*}

\section{Redshifts and Spectral classifications}


\subsection{Redshift completeness and quality flags}

Redshifts and classifications for the objects are shown in Table 3 (full table
in the electronic edition). 
Objects were assigned
a redshift quality flag from 1-4 according to the following criteria: 

\begin{enumerate}

\item Secure redshift based on the high signal-to-noise ($>10\sigma$) detection of two or more spectral features (in emission or absorption).

\item Less secure redshift, based on the detection of one emission line at high signal-to-noise plus
one line at lower signal-to-noise ($>3\sigma$), or more than two lines at lower signal-to-noise.

\item Uncertain redshift based on a single weak line with only weak 
secondary features to obtain the redshift, 
or a single strong line whose identity was deduced from its profile or 
the absence of other 
lines expected to be similarly strong in the spectrum.

\item Featureless spectrum, or spectrum with only a single weak line and no other 
features (no redshift assignment attempted).

\end{enumerate}

\begin{figure*}
\includegraphics[scale=0.6]{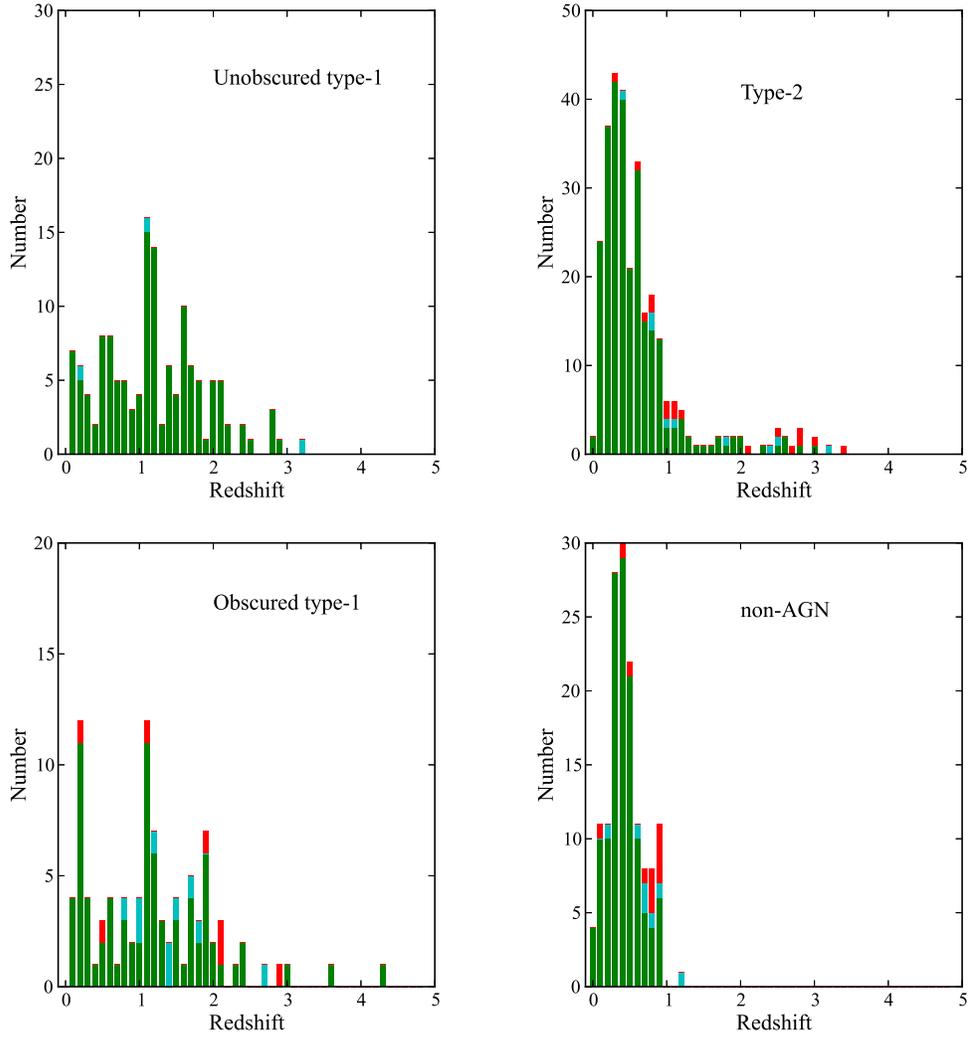}
\caption{Histograms of redshift for each of the four classes of object. Objects
with redshift quality 1 are shown in green, those with redshift quality 2
in blue, and those with redshift quality 3 in red.}\label{fig:zqual}
\end{figure*}

Objects with redshift quality 2 or 3 were also checked for plausibility with the
assumed redshift using SED fitting using models similar to those in Lacy et al.\ (2007b). 
In cases that these led to unphysical SEDs (for example, host galaxy masses $>>10^{12}M_{\odot}$,
or significant dust emission corresponding to dust temperatures above the sublimation
temperature of $\approx 1500$K), the redshifts were assumed to be incorrect and redshift
quality set to 4. Full details of the SED fitting will be given in Lacy et al.\ (2013). 
Figure \ref{fig:zqual} shows the distribution of redshift quality by 
object classification (see Section 4.2 below). As expected, the 
unreddened AGN have the lowest fraction of poor quality redshifts. The
redshift distribution of the non-AGN cuts off abruptly
at $z\approx 1$, probably because higher redshift objects in this 
class lack strong emission features in their rest-frame UV spectra, so 
would have featureless spectra.

Overall redshift completeness in the samples varied from 100\% for most of the 
bright samples, to $\sim 20$\% for some of the faint samples. 
In order to obtain a subset of the survey useful for statistical purposes
in Ridgway et al. (2013), a $\geq$ 90\% complete subsample was computed for each 
individual sample 
by exploiting the 
correlation of $S_{24}$ with emission line flux (except 
for GMOS-S, which has a very narrow range in $S_{24}$). This was carried 
out as follows: each sample was sorted in descending 
order of $S_{24}$. The completeness was then calculated as a function of $S_{24}$ (assuming all redshifts with qualities 1-3 were correct). Objects
were included in the 90\% complete sub-sample until the completeness as a function of $S_{24}$
dropped below 90\% for the final time (see Table 1). These 90\% complete samples, 
combined with the 81\% complete GMOS-S sample constitute the 
``statistical sample'' of 662 objects. 

\begin{figure*}

\includegraphics[scale=0.6]{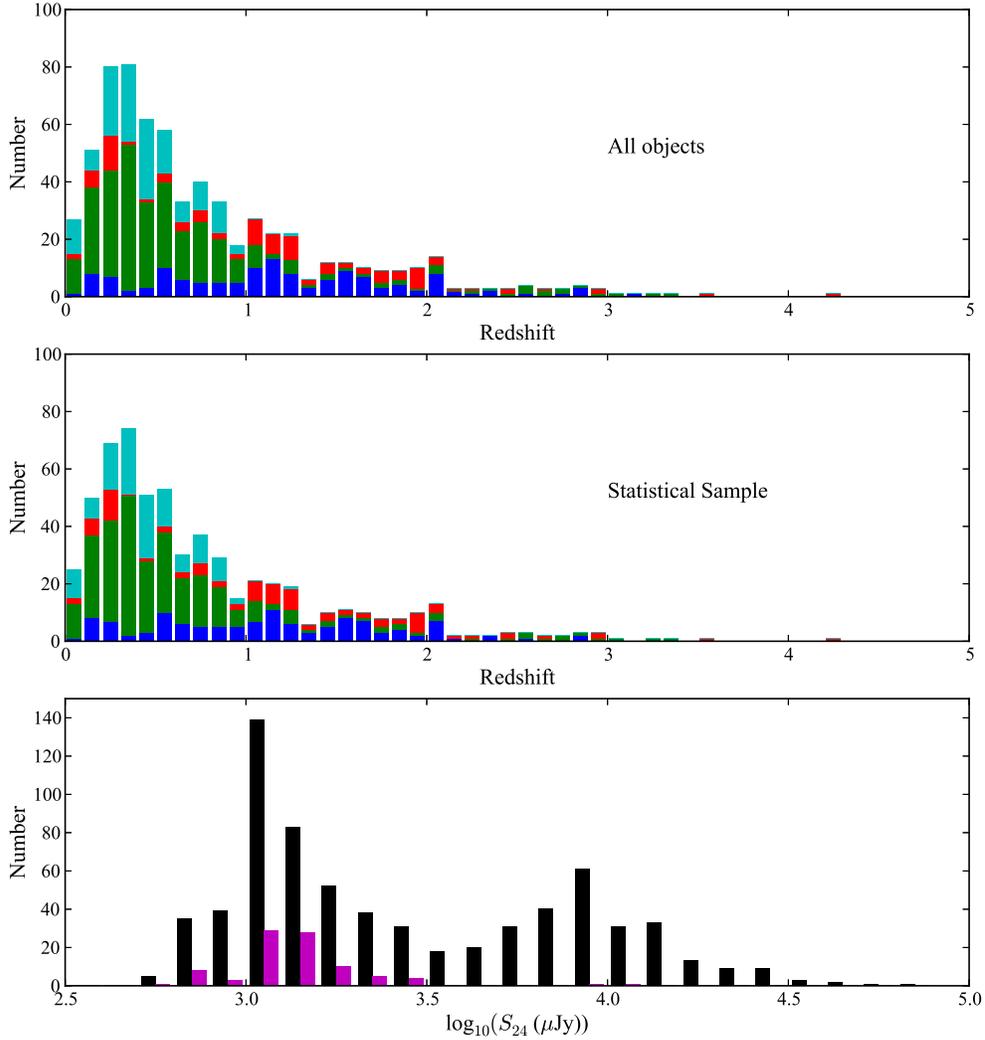}

\caption{Histograms of redshifts by object type: type-1 objects are shown
in blue, type-2 objects in green, reddened type-1 objects in red and objects
with non-AGN optical spectra in cyan. The bottom histogram shows the 24$\mu$m
flux distribution for the objects with featureless spectra (magenta) compared
to the remainder of the sample (black).}\label{fig:zdists}
\end{figure*}

The histogram of redshifts broken up by type (as described below), 
together with the 24$\mu$m flux distribution of the objects lacking redshifts
are shown in Figure \ref{fig:zdists}. As expected, most of the objects lacking
redshifts are faint at 24$\mu$m (see Section \ref{sec:oiii}). There does seem to be a deficit of 
type-2 objects with $1.4<z<1.6$ analogous to the traditional ``redshift desert'' for
normal galaxies, when [OII]3727 redshifts out of the optical band and before
Ly$\alpha$ redshifts in (see also the top-right panel of Figure 7). 
For our type-2 objects, the CIV 1549 line is typically brighter
than Ly$\alpha$, which is we believe why the redshift desert for our objects 
extends only to $z\approx 1.6$, where CIV is shifted to 4000\AA, the blue end
of most of our spectra. The type-1 objects are not affected by this, 
as they have strong low-ionization broad lines (MgII 2798 and CIII]1909)
in the observed optical bands at these redshifts. These low-ionization 
UV lines tend to be relatively weak in the type-2s (see Section 9).
The dip may also be in part due to the fact that 
samples flux limited at 24$\mu$m containing heavily obscured objects 
are biased against these redshifts as the 9.7$\mu$m silicate feature is 
redshifted into the 24$\mu$m band in this range.

\subsection{Classification}

Objects were classified as either type-1 (normal, unobscured; type$=$1 in Table 3),
type-2 (heavily obscured; type$=$2 in Table 3), lightly obscured, with broad lines
visible in the rest-frame optical, but a red continuum (type$=$4 in Table 3), 
or showing no evidence for an AGN in their optical
spectra (type$=$3 in Table \ref{tab:redshifts}, called 
``non-AGN'' in the text of this paper, although some of them in fact 
are AGN based on other criteria, see Section 7). There are also 24
stars, most likely contaminants due to saturated IRAC flux densities (though 
probably possessing debris disks to be bright at 24$\mu$m), 
classified as type$=$5.
The criteria followed those employed by L07, based principally on 
emission line properties (presence of broad lines, classification
on BPT diagrams, presence of [NeV] emission or presence of high 
ionization UV lines in emission), 
although a more refined
definition of dust-reddened versus normal AGN was used, 
as described below.
Where possible, a rest-frame optical spectrum was used to distinguish 
between a type-2 AGN and a lightly obscured AGN.
Objects with uncertain types have type quality
two or greater. Typically these are objects which lack a full set of detected
lines to plot on a diagnostic diagram, or are high redshift without a 
rest-frame optical spectrum. 

We can be fairly sure that 
objects showing narrow, high ionization UV emission lines are AGN based on
comparison with radio galaxies and X-ray detected type-2 quasars (e.g.
Norman et al.\ 2002). Although
high ionization resonant emission lines such as SiIV 1402\AA$\;$and 
CIV1548/1551\AA$\;$can be formed in starbursts, in the
stellar winds and photospheres of the most massive stars, as well as the
interstellar medium (e.g.\ 
Robert, Leitherer and Heckman 1993), they are predicted to form
P-Cygni profiles, with substantial absorption from gas blueshifted along the
line of sight, which we do not see in our objects. Furthermore, in practice,
Lyman break galaxies show high-ionization species predominantely in absorption 
(Shapley et al.\ 2003). High ionization species which do appear in emission,
such as HeII 1640\AA$\;$and CIII] 1909\AA$\;$tend to have very low rest-frame
equivalent widths (RFEWs) ($\stackrel{<}{_{\sim}}2$\AA). We also have several 
known examples of objects which show narrow lines in the rest-frame UV, 
but broad lines in their near-infrared, rest-frame optical spectra 
(e.g.\ XFLS 171053.51+594433.1), which reinforce the point that near-infrared
spectra are needed for accurate classification as a type-2 versus a reddened
type-1 object. For objects which lack near-infrared spectroscopy, we have
been able to make an approximate classification based on SED shapes, however,
these are uncertain and have type quality $=$2 or 3.

A significant fraction of our mid-infrared selected AGN have broad-line 
spectra, but are clearly
redder than the optically-selected quasar population (the objects classified as
type$=$4 in Table 3). This red AGN
population has also been found
in other surveys based on selection from the Two Micron All Sky Survey (2MASS) 
(e.g.\ Cutri et al.\ 2003) 
or 2MASS and the Faint Images of the Radio Sky at Twenty-centimeters (FIRST) 
radio survey (e.g.\ Glikman et al.\ 2005; 2008; Urrutia et al.\ 2009).
Classification of red versus normal AGN needs to take into account 
(at least in a statistical sense) the 
intrinsic variation in AGN colors, and be empirically based, so as
not to be tied to a specific reddening law. We have adopted the technique
of Richards et al.\ (2003), who use the $g-i$ color excess relative to 
the mean for SDSS quasars as a measure of reddening. They argue that intrinsic color variations
are likely to be symmetrically distributed around the mean color, and that the objects with 
red excesses beyond this distribution are likely to be reddened by dust. Most of our objects
have photometry on the SDSS system, and thus we are able to use the $(g-i)$ reference colors
as a function of redshift from Richards et al.\  to calculate the distribution
of excess $(g-i)$ color, $(g-i)^*$, 
for the 193 of our objects which are classified as type-1 (normal or 
reddened) and have photometry on the SDSS system (samples XMM-LSSBright, 
XMM-LSSDeep, LockmanBright, LockmanDeep, EN1Bright, EN1Deep, 
EN2Bright, XFLSBright, XFLSDeep), 
shown in Figure~9(a). The peak around 
zero reddening has a standard deviation (after clipping) of $\approx 0.15$
magnitudes. We therefore place our empirical boundary between reddened 
and unreddened objects two standard deviations away, at 
$(g-i)^*=0.3$. This adoption of an observed-frame color to distinguish red
from normal AGN does mean that the reddening
in the rest-frame at which this occurs will vary as a function of redshift
(as a given optical depth of dust will result in a higher $(g-i)^*$
in high redshift objects),
however, only a handful of objects have colors close enough to the 
boundary to make their classifications uncertain due to this. The 
rest-frame $E(B-V)$ obtained from fitting a reddened 
template quasar SED to SEDs of objects close to our proposed 
boundary implies that a $(g-i)^*$ color excess of 0.3 magnitudes 
corresponds to an $E(B-V)\approx 0.1-0.15$ at $z\sim 1$, previously used
by us to define red quasars (e.g.\ Urrutia et al.\ 2009). The 
24 objects without
colors on the SDSS system (objects in the GMOS-S, CDFSBright, ES1Bright and
ES1Deep samples) were evaluated on the basis of available optical colors as
close to $g$ and $i$ as possible. In most cases classification was unambiguous,
however, in five cases which were
close to the estimated boundary a classification of unreddened was used.

We also used the SED of Richards et al.\ (2006) to calculate the 
$r-[24]$ color excess between the optical and mid-infrared, $(r-[24])^*$
(Figure~9(b)). There is a clump of normal AGN near $(0,0)$, 
consistent with the idea that we are dominated by the normal AGN 
population, and are not systematically selecting 
objects with unusually bright infrared emisson. 
As expected for dust reddening, $(g-i)^*$ correlates
broadly with $(r-[24])^*$, though the correlation is surprisingly 
poor. The objects scattered towards bluer 
$(r-[24])^*$ colors seem to be explained in terms of host galaxy 
contamination, as the majority of objects in this area of 
the plot are of low redshift and
show evidence of host galaxy starlight in their spectra. Conversely, 
the outliers with high $(r-[24])^*$ and low $(g-i)^*$ tend to be 
objects very faint in the optical, where again host galaxy emission and
(possibly) scattered AGN light may make the observed colors bluer. The lack 
of a tight correlation for the more highly 
reddened objects (redder than $(g-i)^*=0.3$) 
may also reflect genuine variation in the ratio of reddening
(measured by $(g-i)^*$) to optical extinction (approximated by $(r-[24])^*$)
in our population.

\begin{figure*}
\begin{picture}(400,150)
\put(0,0){\includegraphics[scale=0.45]{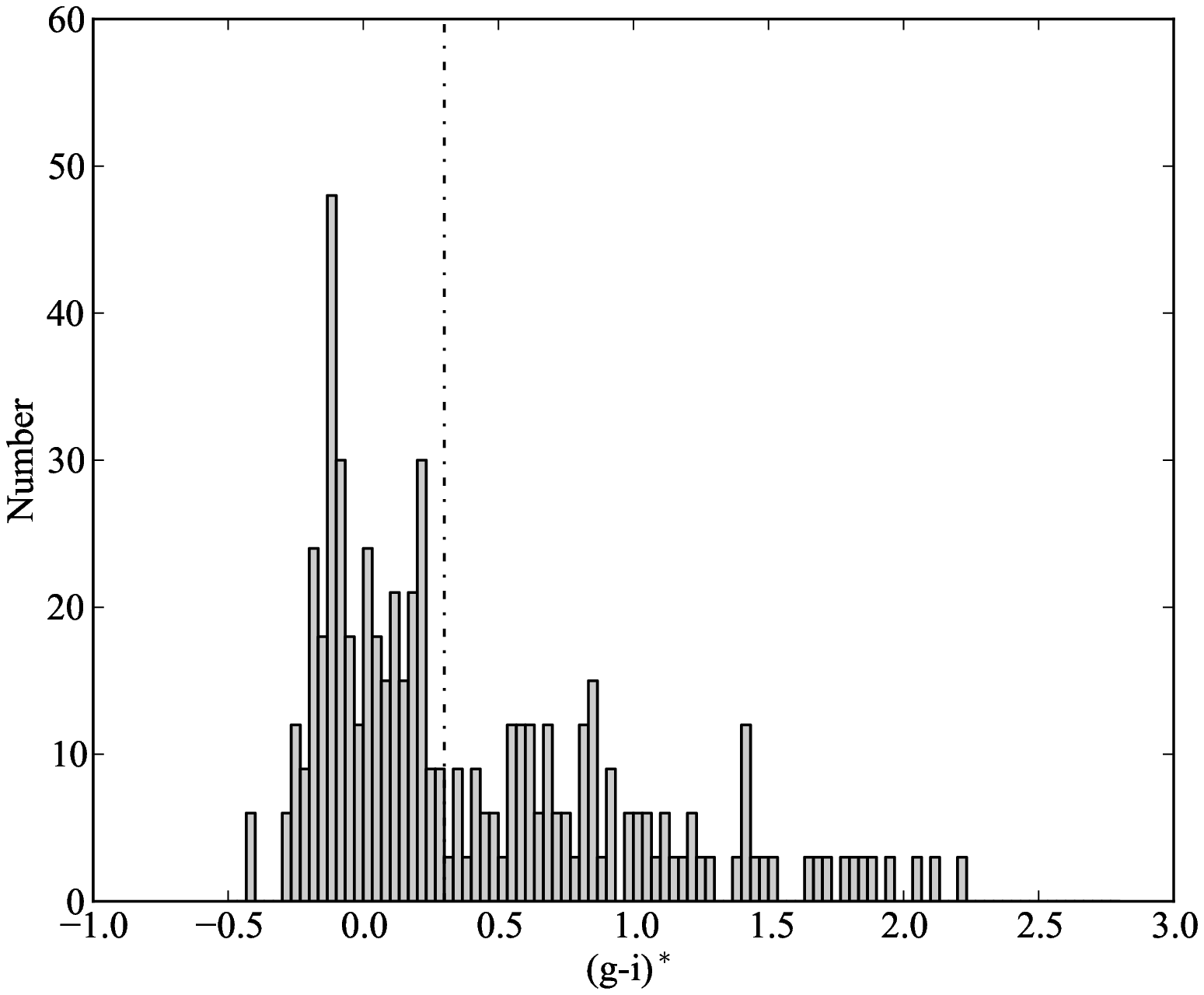}}
\put(230,0){\includegraphics[scale=0.45]{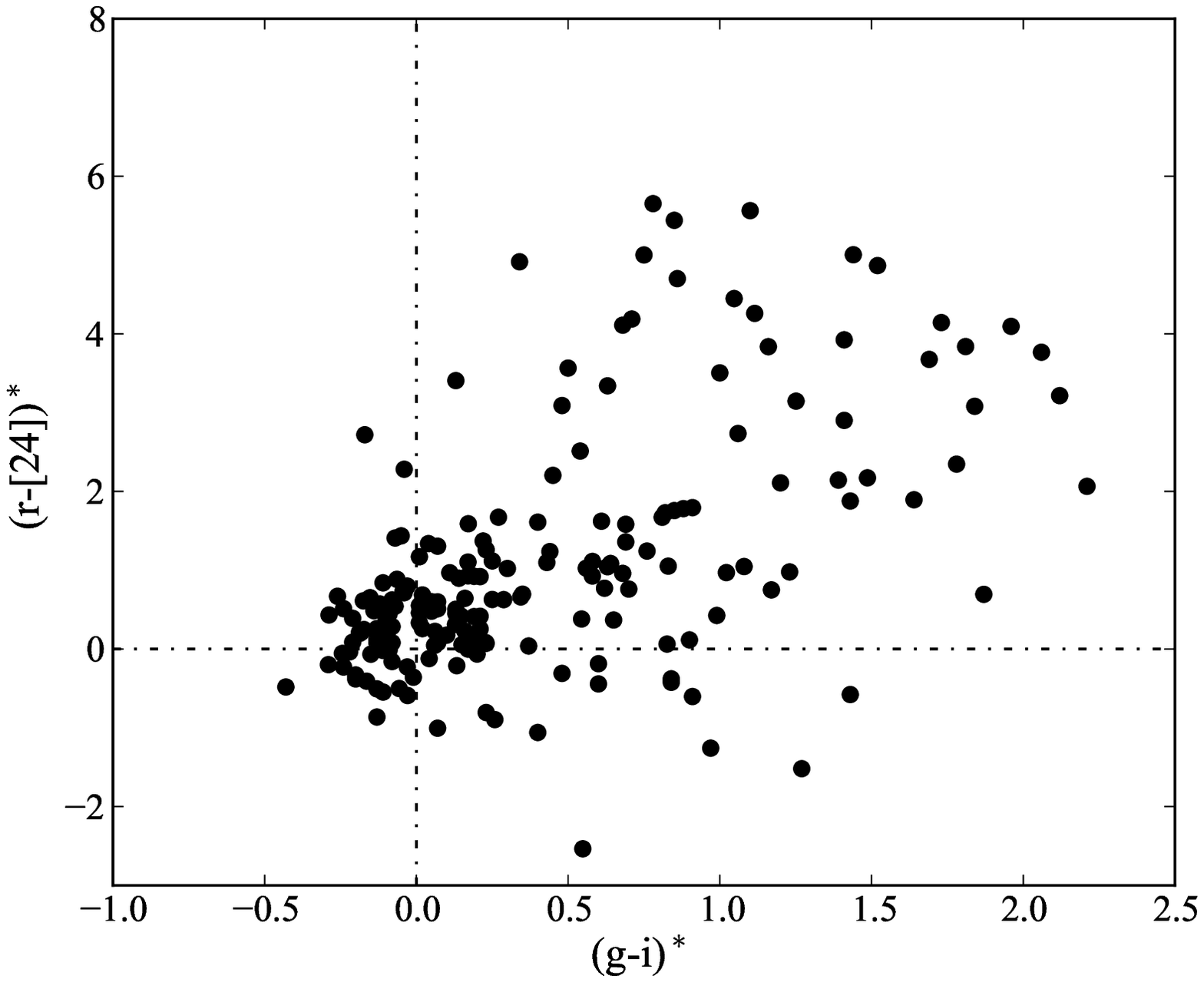}}
\end{picture}
\caption{{\em Left} (a) histogram of corrected $g-i$ color for the objects 
classified as type-1 (i.e. having broad lines in the rest-frame optical). 
{\em Right} (b) excess $g-i$ color plotted against corrected $r-[24]$ color
for the broad-line objects.}

\end{figure*}



\section{X-ray detections}
\label{sec:xray}

We matched the objects in our spectroscopic survey in Table 3 
to the 2XMM-DR3 catalog of serendipitous XMM sources
(Watson et al.\ 2009) and the {\em Chandra} Source Catalog (CSC; Evans et al.\ 
2010), finding 108
matches to 2XMM and 81 to the CSC (36 of which are in both catalogs)
(Table 5). Both catalogs have only patchy
coverage across the survey fields, and exposure times
ranging from $\sim 5-50$ks, 
so the low detection
fraction is no surprise. As pointed out in L07 and Donley et al.\ (2012), 
though, many obscured AGN are not detected
even in deep X-ray surveys. The breakdown, among the X-ray detected
sources, by type is 52 normal type-1s, 
45 type-2s, 31 red type-1s, nine objects with redshifts but no
optical AGN signatures, 
14 objects with featureless spectra and two stars. Four of the 2XMM sources
and one of the CSC sources are weaker than $10^{42} {\rm ergs^{-1}cm^{-2}}$,
where their X-ray luminosity could plausibly be powered by X-ray 
binaries in the host galaxy (e.g.\ Teng et al.\ 2005), but the vast majority are much more luminous, and almost
certainly powered by AGN.

\begin{sidewaystable}

\caption{X-ray properties of AGN in the spectroscopic survey}

{\scriptsize
\begin{tabular}{lccclcclcc}

Name & Redshift & Type & $S_{24}$ & CSC Name & CSC 2-10keV & CSC HR & 2XMM Name & 2XMM 0.2-12keV & 2XMM HR2\\
     &          &      & ($\mu$Jy)    &          & flux        & (h-s)  &           & flux           &         \\
     &          &      &          &          &(ergs$^{-1}$cm$^{-2}$)&&           &(ergs$^{-1}$cm$^{-2}$)&\\\hline
SW003312.39-431841.9&0.248&1&1030.0&CXO J003312.3-431842&2.61E-14&0.716161&2XMM J003312.3-431841&2.60E-14&0.93102986\\
SW003316.92-431706.3&&&2200.0&CXO J003316.9-431706&8.12E-15&0.375333&2XMM J003317.0-431705&4.29E-14&0.917\\
SW003330.42-431554.3&0.437&3&1890.0&CXO J003330.4-431554&3.77E-15&-0.325&2XMM J003330.4-431552&5.79E-15&1.0\\
SW003333.75-432326.9&&&1490.0&CXO J003333.7-432326&4.94E-15&0.0643&2XMM J003333.7-432328&1.98E-14&0.651\\
SW003336.26-431731.7&&&2190.0&CXO J003336.2-431731&1.82E-14&0.482&2XMM J003336.1-431732&3.58E-14&0.837\\
SW003346.28-431943.7&0.402&2&3240.0&CXO J003346.2-431943&4.02E-14&0.025&2XMM J003346.2-431944&6.40E-14&0.176\\
SW003357.32-433601.7&1.932&4&1060.0&CXO J003357.2-433602&1.24E-14&-0.341&&&\\
\end{tabular}

Notes:- This table contains the matches in the CSC and 2XMM catalogs to all the objects in Table 3 (which 
have with either spectroscopy from the literature or from our own observations). Objects with featureless
spectra are included (with the redshift and type columns left blank).
Table 5 is published in its entirety in the electrononic edition of Astrophysical Journal Supplement, a portion is shown here for guidance regarding its form and content. 
}
\end{sidewaystable}

\begin{figure*}

\includegraphics[scale=0.6]{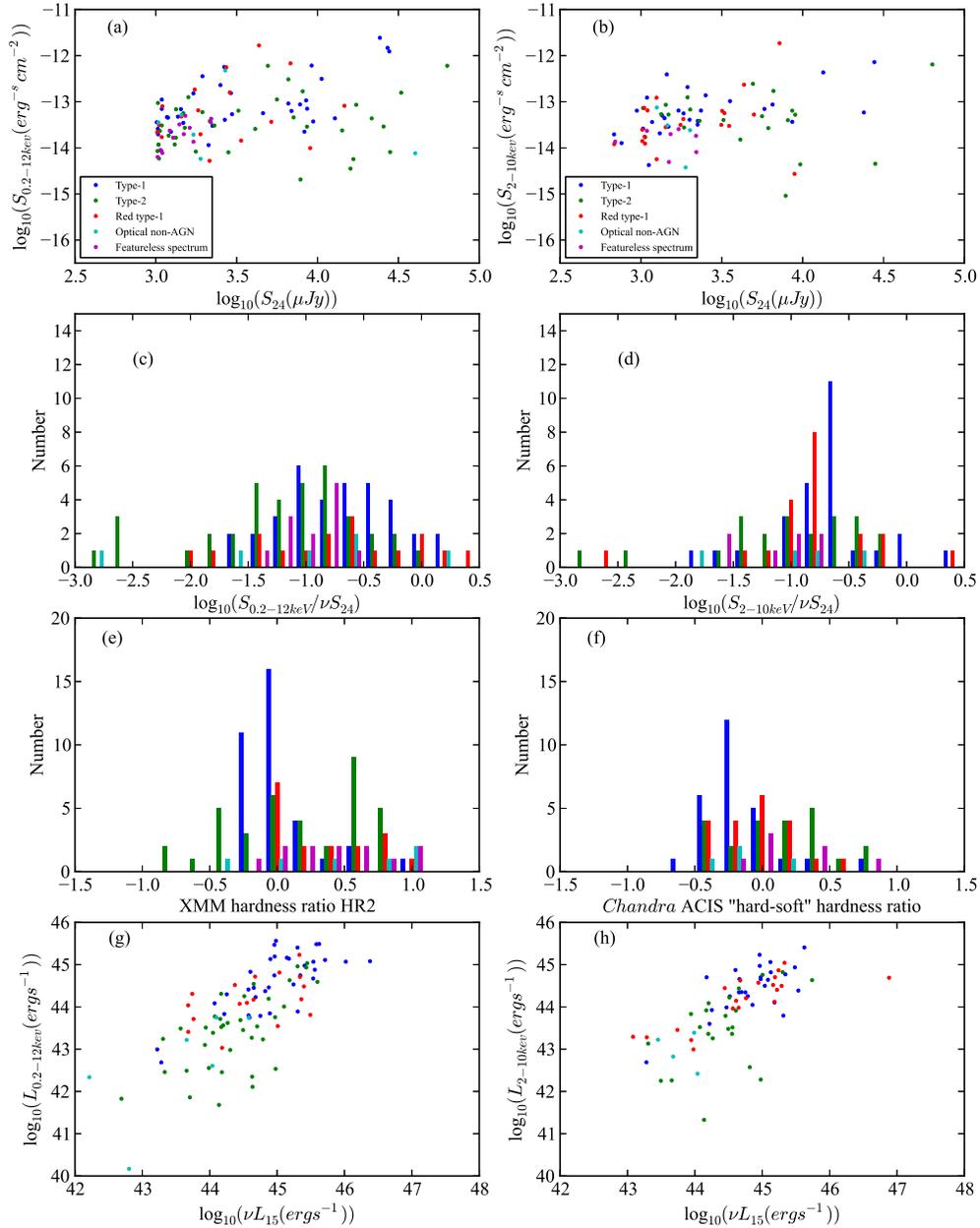}

\caption{X-ray detections and hardness ratios of objects in the survey by type. 
(a) XMM 0.2-12keV flux versus 24$\mu$m flux density, 
(b) {\em Chandra} 2-10keV flux versus 24$\mu$m flux density, (c) ratio of XMM 0.2-12keV to 24$\mu$m flux, (d)
ratio of {\em Chandra} 2-10keV flux to 24$\mu$m flux,
(e) distribution of XMM hardness ratio HR2, (f)
distribution of $Chandra$ hardness ratio, (g) XMM luminosity versus 15$\mu$m luminosity, and (h) $Chandra$ luminosity versus 15$\mu$m luminosity. The color coding of the histograms is given in the top panels.}\label{fig:xray}

\end{figure*}

Figure \ref{fig:xray} plots the X-ray flux against the 24$\mu$m flux, the
histogram of the X-ray to 24$\mu$m flux ratio and 
the distribution of hardness ratio (HR) by type (for XMM we use
$HR2$, defined to be that between 0.5-1keV and 1-2keV, for Chandra we 
use the HR defined between the ``hard'' and ``soft'' ACIS 
bands, called $HRC$ in this paper), and the X-ray luminosity versus the mid-infrared luminosity. 
The mid-infrared k-correction is calculated using the measured 8-24$\mu$m 
spectral index. 
For both the 2XMM and CSC detections, the type-1
objects are generally the brightest and softest in X-rays
(mean $HRC$, $<HRC>=-0.13$ with a dispersion of $\pm 0.25$, mean $HR2$, $<HR2>=0.02\pm0.25$) and 
their X-ray fluxes correlate loosely with the 24$\mu$m flux. The 
type-2s and other obscured populations are, on average, the faintest and 
have a wide range of hardness ratios ($<HRC>=0.14\pm 0.37$; 
$<HR2>=0.15 \pm 0.48$ for
Type-2s, $<HRC>=-0.04\pm 0.31$; $<HR2>=0.31 \pm 0.35$ for red type-1s, 
$<HRC>=-0.14\pm 0.22$; $<HR2>=0.45 \pm 0.53$ for optically-classified ``non-AGN'', 
and $<HRC>=0.21\pm 0.34$; $<HR2>=0.40 \pm 0.37$ for objects
with featureless optical spectra. 
If we were to reclassify objects purely on the basis
of their X-ray hardness ratios, and assumed a division of $HRC= 0$ or $HR2=0.1$
as the boundary of ``obscured'' versus ``unobscured'', 19\% (12/62) of the optically-classified
unobscured type-1 objects would be classed as obscured in the X-ray. 
Conversely, 41\% (24/59) of the type-2 objects would have been classified as X-ray
unobscured, along with 51\% (19/37) of the reddened type-1 objects, 
50\% (5/10) of the objects with non-AGN spectra, and 26\% (7/12) of the 
objects with featureless spectra. These differences are, however,
subject to systematic uncertainties as we only consider X-ray detected
objects, and the obscured objects in particular tend to be faint, 
rendering their hardness ratios uncertain, and subject to 
contamination by soft emission from scattering of AGN 
emission, or X-ray binaries in the hosts. Nevertheless
our results are consistent with those of Brightman \& Nandra (2011), who
also note discrepancies between X-ray and optical classifications of AGN.
On the X-ray luminosity -- mid-infrared luminosity 
plots the correlation for the 2XMM points
(whose fluxes are more weighted 
towards the soft X-rays) is less tight
than for the 2-10keV $Chandra$ luminosities. 

\section{Radio detections}
\label{sec:radio}

The fields used for this survey overlap with several moderate depth 
(root mean square (RMS) sensitivity $\approx$10-30$\mu$Jy) 
radio surveys at 1.4GHz. In the north, the XFLSBright and XFLSDeep 
samples overlap with the Very Large Array (VLA) survey of Condon 
et al.\ (2003), the LockmanDeep sample 
overlaps with the Ibar et al.\ (2009) VLA survey and the ELAIS-N1Deep sample
overlaps with the VLA survey of Ciliegi et al.\ (1999). In the south, the
ELAIS-S1Bright, ELAIS-S1Deep and CDFSBright surveys overlap the Deep 
Australia Telescope Large Area Survey made with the Australia
Telescope Compact Array (ACTA) (Norris et al.\ 2006; 
Middelberg et al.\ 2008). We have cross-matched our survey AGN 
with the public source lists from these surveys in order to make an 
initial examination of the radio properties of our objects, detecting
214 objects. In the 
XFLSBright and XFLSDeep fields, which overlap
the deep (RMS 10$\mu$Jy) 
and uniform radio survey of Condon et al., our detection rate is 
90/223, or 40\%, so even in the fields with the best radio data 
the majority of objects remain upper limits. 

\begin{figure*}
\includegraphics[scale=0.5]{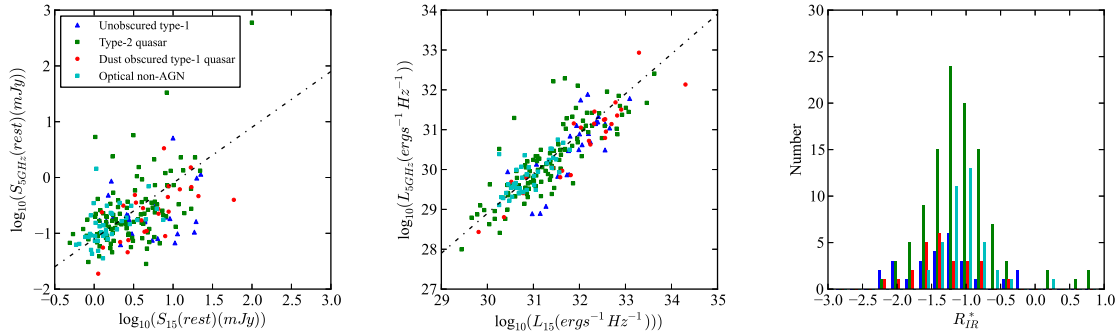}
\caption{{\em Left} rest-frame 5GHz flux density plotted against
rest-frame 15$\mu$m flux density, {\em middle} 5GHz luminosity as a function
of 15$\mu$m luminosity, and {\em right} a histogram of radio-loudness, split
by object type. The dot-dashed lines on the left and middle plots show 
the position of the far-infrared -- radio correlation from Appleton 
et al.\ (2004), scaled to 15$\mu$m and 5GHz, assuming
a mid-infrared spectrum flat in $\nu f_{\nu}$ and a radio
spectral index of -0.8.}\label{fig:radio}
\end{figure*}

Figure \ref{fig:radio} shows the flux-flux plot, the luminosity-luminosity 
plot and a histogram of radio loudness (where we define the k-corrected
infrared radio-loudness as:
\[R^*_{IR}={\rm log_{10}}({F_{5GHz}/F_{15\mu m}}) \] 
where $F_{5GHz}$ and $F_{15\mu m}$ are the rest-frame fluxes at 
5GHz and 15$\mu$m, respectively). A radio
spectral index $\alpha=-0.8$ (where flux density $S_{\nu}\propto \nu^{\alpha}$)
was assumed.
We also show the approximate position of the far-infrared -- radio
correlation adapted from Appleton et al.\ (2004) with a small (-0.2 dex) 
correction to allow for the different infrared and radio wavelengths used.

Most AGN lie close to the far-infrared -- radio correlation, even when the 
mid-infrared, which is dominated by AGN luminosity, is used, as has been
noted in the past (Sopp \& Alexander 1991; Kimball et al.\ 2011). We also see
no sign of any radio-loudness dichotomy or bimodality in any of our object
classifications, consistent with most recent studies (White et al.\ 2007; 
Kimball et al.\ 2011).  We do, however, see significant 
scatter both above the far-infrared -- radio relation 
(the expected radio-loud/intermediate population) and below (where
objects with strong AGN-related mid-infrared emission appear). 
Furthermore, the
distribution of $R^*_{IR}$ values for the broad-line objects (with a mean
of -1.35 with an error in the mean of $\approx 0.1$ 
for both the normal and obscured type-1 
populations) is lower than
that for the type-2 population ($-1.09\pm 0.05$) and the non-AGN 
($-1.05\pm 0.05$). We performed both a Mann-Whitney $U$-test and a $t$-test to 
investigate whether the values of $R^*_{IR}$ are systematically lower for the 
broad-line objects taken as a group versus the type-2s, finding a 3\% 
probability of the samples being drawn from identical parent populations
in both cases. Thus the difference 
is not highly robust, and, furthermore, the large number of upper limits
may mean that the result may be biased by the different redshift and 
luminosity 
distribution of the broad line and type-2 populations (although there is
no obvious luminosity dependence in Figure \ref{fig:radio}). 
If this result is real, however, one interpretation
can be obtained if we assume that the 
intrinsic mean ratio of mid-infrared to radio emission is the same
for both populations, but that the type-2s have some extinction towards the 
mid-infrared, even at 15$\mu$m. This is broadly consistent with the work 
on radio-loud AGN by Cleary et al.\ (2007) and Haas et al. (2008), where 
the infrared SEDs of radio galaxies and radio-loud quasars matched
in radio luminosity differ by a factor $\approx 3$ at $12 \mu$m (i.e.\ 
very similar to our $\approx$ factor of two difference at 15$\mu$m), 
despite having similar total FIR luminosities, and may 
also be supported by our tentative findings 
in Section 8 regarding the optical depth of the torus. However,
this must remain speculative pending better radio data and consideration
of the full infrared SEDs and demographics of these objects.

\section{The nature of objects lacking AGN signatures in the optical}
\label{sec:type3}

A significant fraction (22\%) of our objects 
with redshifts show no clear AGN signatures in
their optical spectra and are classified as ``non-AGN''. 
Where spectral features are seen, these
objects have low ionization emission line 
spectra. An additional 15\% of our candidates have
featureless spectra or spectra with only a single, unidentifiable 
faint line (redshift quality$=$4).
They are excluded from our AGN sample, but may contain AGN with both 
the NLR and AGN continuum/broad-line region heavily obscured. 
In Figure \ref{fig:wedge} they are spread throughout the selection region, 
some close to the edges, suggesting that indeed they are interlopers, 
but some are directly along the AGN locus, and satisfy 
the Donley et al.\ (2012) selection criteria.
In a few of the objects we have ancillary information
indicating an AGN. 
Nine objects with ``non-AGN'' spectra are detected 
in the 2XMM or CSC catalogs, including SW021822.13-050614.1, which Severgnini
et al.\ (2003) show has faint AGN emission lines in its nuclear spectrum
once a host galaxy template is subtracted from a deep optical spectrum. 
14 objects with optically-featureless or single emission line
spectra are detected in X-rays. 
Finally, one object is radio-intermediate, with $R^*_{IR}=0.13$
(SW160857.99+541818.4) and one object,  XFLS171115.2+594906,
has a flat radio spectrum, indicative of its radio emission arising from an AGN
(a spectral index between 610MHz and 1.4GHz
of $\alpha=-0.3$, where radio flux density $S_{\nu}\propto \nu^{\alpha}$).

Based on comparing the X-ray detection rate of type-2 AGN and 
that of objects lacking AGN signatures we
can place a rough lower bound on the number of AGN missing in our 
optically-classified sample. 45 optically-classified 
type-2 AGN are detected in X-rays, out
of a total of 295 in the survey, a detection rate of 15\%. This compares to 
23 objects with featureless spectra or otherwise 
no sign of AGN in their optical spectra, out
of a total of 233 such objects in the survey, a detection rate of 10\%, only a 
little lower than that of the type-2s. If we assume that 
the ratio of X-ray detected to X-ray undetected AGN is similar in the
objects optically classified as ``non-AGN''  as it is in the type-2s 
this implies that at 
least $\approx 65$\% of the objects classified as ``non-AGN'' 
on the basis of their optical spectra may in fact 
contain one.

\section{The emission-line - mid-infrared luminosity correlation}
\label{sec:oiii}

\begin{figure*}
\includegraphics[scale=0.5]{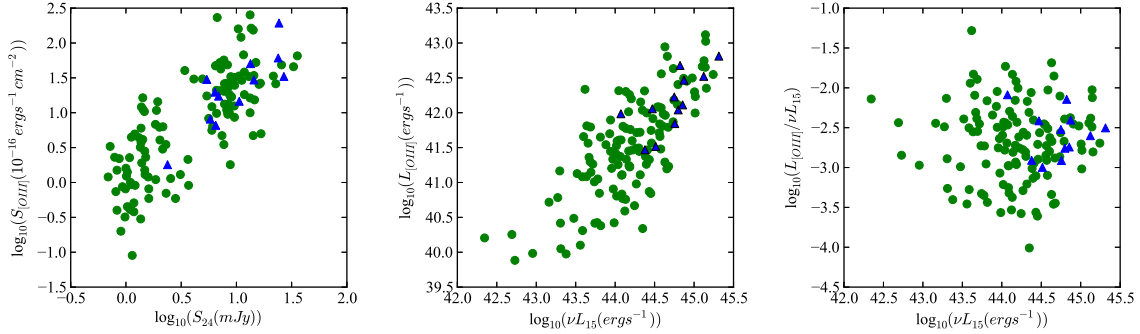}
\caption{{\em Left} log of [OIII]5007 flux, $S_{[OIII]}$ plotted against
$S_{24}$. {\em Middle} log of [OIII]5007 luminosity, $L_{\sc [OIII]}$ plotted
against 15$\mu$m luminosity ($\nu L_{15}$). {\em Right} log of the ratio
of [OIII] to 15$\mu$m luminosity against 15$\mu$m luminosity. In all plots, 
type-2 AGN are plotted as green dots and type-1 AGN as blue triangles.}\label{fig:oiii} 
\end{figure*}

Both the higher ionization emission lines and the mid-infrared emission are powered by 
UV emission from the AGN, so we expect these two quantities to correlate well.
To test this, we measured the [OIII]5007 flux in normal type-1 and
type-2 AGN spectra at $z<0.75$, where this line is present in the 
optical spectra. As shown in 
Figure \ref{fig:oiii}, the correlation exists between both fluxes and luminosities, with
about a half dex scatter. The origin of this scatter is unclear. Although
we see evidence of dust reddening in the type-2 composite (see Section 9), the
fact that the type-1 AGN plot among the type-2s and with similar scatter
suggests that reddening is not the primary cause of the scatter. 
Other possibilities 
include variation in ionization level of the narrow-line region (NLR) 
(unlikely given the 
similarity of the NLR spectra we see), or variation in the 
covering factor of dust in the nuclear region. This latter explanation seems
the most likely, though it is surprising that there is no strong correlation
of this ratio with luminosity. Such a correlation would naturally occur in the simplest version
of the ``receding torus'' model (Lawrence 1991; Simpson 2006), where 
(in the case of a torus optically thin in the mid-infrared) the
mid-infrared luminosity should scale proportional to the covering factor of 
hot dust, $\omega$, and the [OIII] luminosity should scale proportional
to $(1-\omega)$. The large scatter means that our result is probably
consistent with the relatively weak dependence
of covering factor with AGN luminosity found by Roseboom et al.\ (2012),
though may be inconsistent with the optically-thin torus case of Lusso et al.\
(2013) (and indeed Drouart et al.\ (2012) show that an optically-thick
torus is a good fit to radio galaxy SEDs when using the radio core-to-lobe
flux ratio to constrain the inclination of the torus). 

\section{Composite spectra}
\label{sec:comp}

We constructed a composite spectrum for type-2 quasars, and also for the
``non-AGN'' class to search for evidence of weak AGN activity. Attempts
to construct composites for the normal and reddened quasars did not
result in useful spectra due to strong continua, which (particularly 
for the reddened quasars) varied considerably from object to object.
The type-2 quasar composite was constructed using 
spectra from the bright samples 
(limiting fluxes $S_{24}>4$mJy) 
only at $z<0.8$ in order to exclude objects with Seyfert-like 
luminosities. At $z>0.8$, all samples contained objects of quasar-like 
luminosity, so all available spectra were combined. We used
only spectra from our observations 
in Table 2 that were of good quality (S/N on brightest line
$\stackrel{>}{_{\sim}}30$), and which had good sky subraction.
Two of the fiber spectra, SW160929.35+542940.8 and SW160828.55+542546 had
a low-order continuum subraction performed before addition as they had bad
baselines due to poor sky subtraction.
 
The composite spectrum at wavelength $\lambda_i$ was constructed as:
\[ F(\lambda_i) = \frac{1}{\sum\limits_j (w_{j})}\sum\limits_j w_{j} a_j f_j(\lambda_i) \]
where the sums are over all the spectra contributing at wavelength $\lambda_i$. 
The emission line flux scales approximately with the 24$\mu$m flux 
density $S_{24}$ (see Figure \ref{fig:oiii}), so the scale factor 
$a_j \propto 1/S_{24}^j$. The typical signal-to-noise, however, scales 
proportional to $S_{24}$ 
(resulting in weights $w_j \propto S_{24}^j$).

\begin{figure}

\plotone{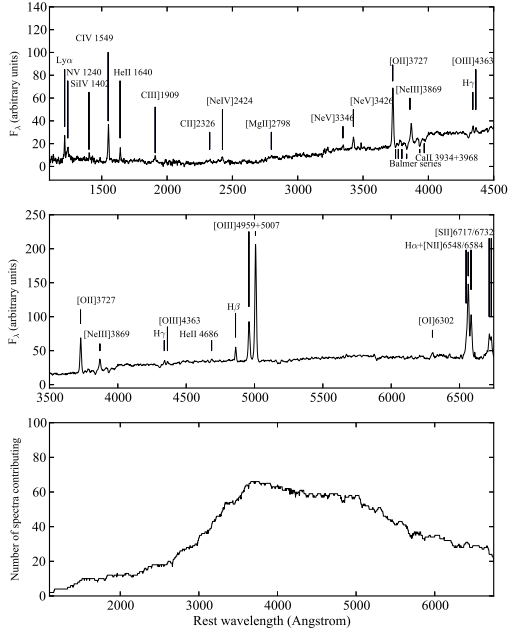}

\caption{Type-2 quasar composite spectrum. The bottom panel shows the
number of quasar spectra contributing at each wavelength.}\label{fig:t2comp}

\end{figure}

The type-2 quasar spectrum (Figure \ref{fig:t2comp}) 
looks very similar to Seyfert-2 
spectra, with strong high-ionization narrow lines, but also including 
low-ionization species such as [OI], consistent with
the broad range of ionization expected from
an AGN (though lacking the coronal lines seen in the 
Rose et al.\ (2011) spectrum of SDSS J113111.05+162739.5). 
The composite host galaxy shows 
multiple Balmer absorption features, consistent with a dominant stellar 
population $\stackrel{<}{_{\sim}}10^8$yr old, as has been seen in the hosts
of IR-luminous type-1 quasar hosts (Canalizo \& Stockton 2000). 

\begin{table}
\caption{Line ratios in composite spectra relative to H$\beta$}
{\scriptsize
\begin{tabular}{lccccc}
Line & Wavelength & Type-2       &Non-AGN& Radio      & Case B\\
     &            &              &       & Galaxy$^*$ &       \\
     & (Angstrom) & $f$/H$\beta$ &$f$/H$\beta$&$f$/H$\beta$&$f$/H$\beta$\\\hline
Ly$\alpha$& 1216  & 0.75          &   -     & 31 &31\\
NV   &  1240      & 0.93         &   -     & 1.5&-\\
SiIV & 1402       & 0.21          &   -     & 1.6&-\\
CIV  & 1549       & 1.45          &   -     & 3.6&-\\
HeII & 1640       & 0.40          &   -     & 3.2&-\\
CIII]& 1909       & 0.34         &   -     & 1.8&-\\
CII] & 2326       &  0.16      &   -     & 0.92&-\\
\[NeIV]& 2424      & 0.40         &   -     & 0.90&-\\
MgII & 2798       & 0.75         &   -     & 0.78&-\\ 
\[NeV]& 3346       & 0.16         &   -     & 0.20&-\\
\[NeV]& 3426       & 0.53         &   $<$0.1& 0.69&-\\
\[OII]& 3727       & 2.78          &  2.5    & 3.64&-\\
\[NeIII]& 3869     & 0.93          &  0.33   & 0.82&-\\
H$\gamma$ & 4340  & 0.19         &   -     & 0.24&0.47\\
\[OIII] & 4363     & 0.18         &  $<$0.25 & 0.08&-\\
HeII   & 4686     & 0.18         &   -      & 0.20&-\\
H$\beta$&4861     & 1.0          &   1.0   & 1.0  &1.0\\
\[OIII] &4959      & 3.9          &   0.38  & 3.1  &-\\
\[OIII] &5007      & 10.8         &   0.99    & 8.7&- \\
\[OI]   &6302      & 0.41         &    -    &  -   &-\\
\[NII]  &6548      & 1.4          &    -     &-    &-\\
H$\alpha$ &6563   &  7.0          &    -     &-    &2.9\\
\[NII]  & 6584     & 4.2          &    -     &-    &-\\
\[SII]  & 6717     & 2.3          &    -     &-    &-\\
\[SII]  & 6732     & 2.0          &    -     &-    &-\\\hline
\end{tabular}

\noindent
$^*$McCarthy (1993)

}\label{tab:comp}
\end{table}

\begin{figure}

\plotone{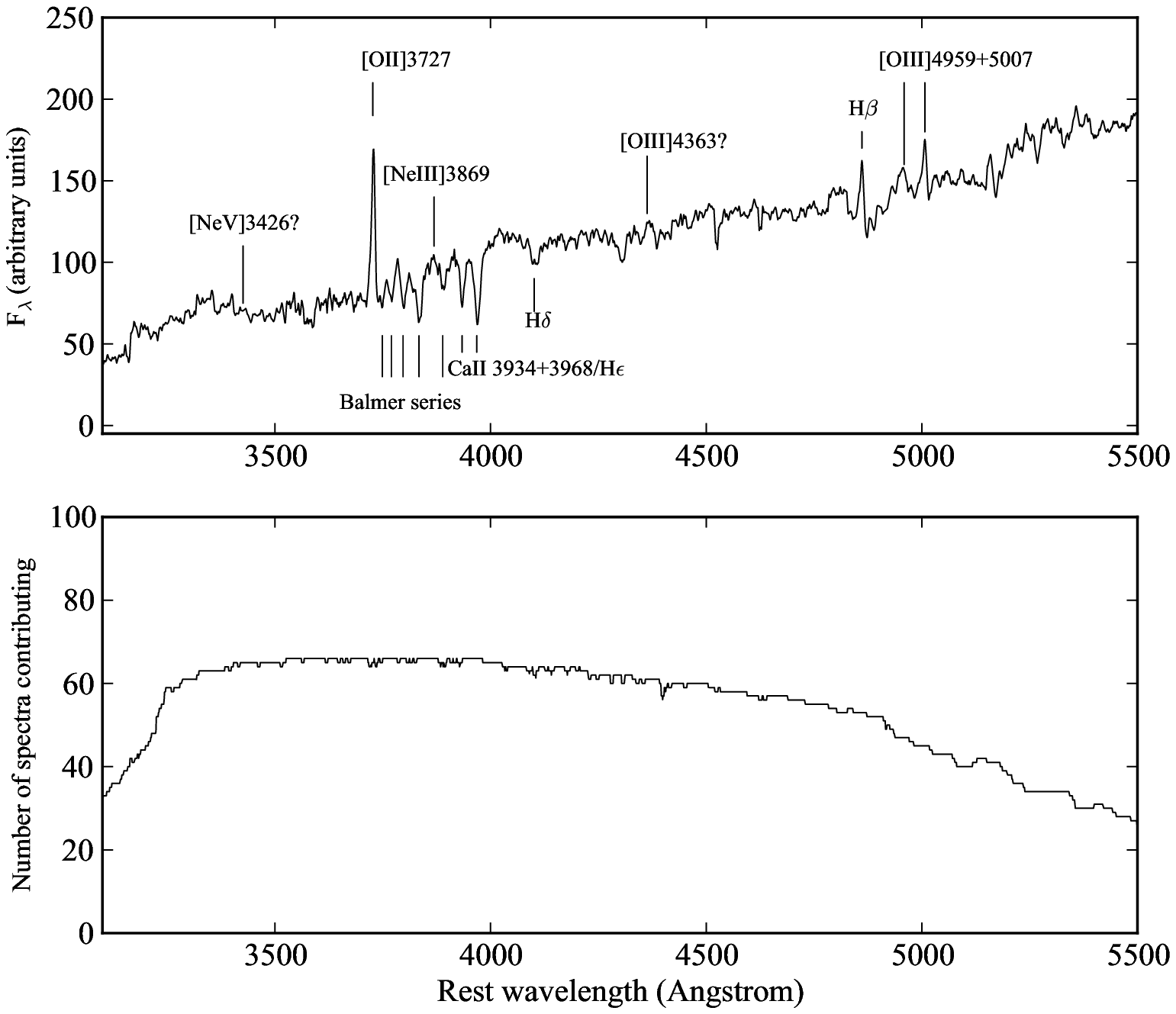}

\caption{The composite spectrum of objects lacking AGN signatures in their 
individual spectra. The bottom panel shows the
number of spectra contributing at each wavelength.}\label{fig:t3comp}

\end{figure}

The ``non-AGN'' spectrum
(Figure \ref{fig:t3comp})  was constructed from an average
of all the spectra for the objects lacking AGN features in their optical
spectra, it is restricted to the optical
as there are no examples at $z>1$ (at least not ones with detectable
emission lines). The emission lines in both the individual spectra and
the composite are 
more characteristic of starbursts than LINERs, as noted by L07, the number
of candidate LINERs in these mid-infrared selected samples is very small, and
we have no definitive LINER candidates.
We have carefully examined our composite spectrum for signs of 
high ionization lines that could indicate the presence of a weak or 
highly obscured AGN but fail to find
any conclusive evidence of these. Perhaps the best evidence from the 
composite that these objects are
similar to the type-2s is the very similar host galaxy spectra, including
Balmer absorption lines of similar equivalent width at H9 and higher. These
objects could thus be obscured AGN with significant reddening 
towards the NLR, or simply lacking one. Nevertheless, we choose to 
exclude them from this AGN census until more unequivocal evidence for 
their AGN is found.

Table \ref{tab:comp} compares the emission line fluxes from our Type-2 and non-AGN
composites with the radio galaxy template from McCarthy (1993), and 
a Case B recombination model (for the hydrogen lines). 
Our type-2 composite spectrum 
shows evidence for reddening, in terms of significant Balmer 
decrements between H$\alpha$, H$\beta$ and H$\gamma$, and a weak Ly$\alpha$
line. There are also differences between the type-2 composite
and that of the radio galaxies, in particular, the low-ionization UV lines
(e.g. CIII] and CII] are weak compared to CIV in our composite when compared 
to radio galaxies) and the Ly$\alpha$ line is much weaker
in the type-2 spectrum than in the radio galaxy composite. We speculate that 
this may be due to the effect of jet-induced shocks in the radio galaxies
(a result consistent with the morphologies of extended emission line nebulae of
type-2 quasars studied by Liu et al.\ 2013), 
but defer a detailed analysis to a future paper.

\section{Discussion}

Our survey of 786 objects with (attempted) spectra contains 672 
extragalactic objects with 
redshifts (24 objects are stars included in error due to saturated 
flux densities in the IRAC bands, and 90 objects had featureless spectra).
Of these, we have classified 136 as type-1 
(normal quasars/Seyfert-1s), 96 as type-1 objects showing significant
signs of reddening, 294 as type-2 quasars/Seyfert-2s, leaving 
145 objects with redshifts, but showing no sign of AGN activity in their 
optical spectra. Our nested survey strategy has
enabled us to span a wide range in luminosity ($>$ a factor of ten) 
at a given redshift. We classify 340 (50\%) of the
survey AGN as quasars based on their mid-infrared
luminosities indicating that their bolometric accretion luminosities are
$\stackrel{>}{_{\sim}}10^{12}L_{\odot}$.
The difference in extinction in the mid-infrared
between type-1 and type-2 objects is probably at most a factor of $\approx 3$
(see discussion in Section \ref{sec:radio}). Thus we are able to select
very comparable objects with both normal type-1 and obscured quasar 
natures in the rest-frame optical.

We defined a ``statistical survey'' with 90\% complete redshift
information in order to allow for the inevitable 
bias towards type-1 objects in our spectroscopic
survey. Members of this survey were 
identified by ordering each sample contributing to the 
survey in 24$\mu$m flux density high to low and including objects 
in the statistical survey by working down the list until
the spectroscopic completeness fell below 90\% (if at all). 
This sample consists of 662 objects, with 122 normal type-1s, 86 reddened 
type-1s, 271 type-2s and 118 ``non-AGN''. 23 objects were stars and 42
objects lacked redshifts. Obscured AGN thus dominate the population 
over the redshift and luminosity range probed by this survey.
We will use this survey to discuss the demographics and evolution 
of these objects in a future paper (Ridgway et al. 2013). 

The composite spectrum of the type-2 quasars shows high ionization narrow
lines, but is different in the ultraviolet to radio galaxies,
with weaker Ly$\alpha$ and other
low-ionization species. The [OIII]5007 line emission of our 
objects correlates
well with their 24$\mu$m luminosity, and both type-1 and type-2 objects lie
on the same correlation, showing that mid-infrared selection is indeed
capable of selecting type-1 and type-2 quasars with similar intrinsic 
properties, with no apparent strong dependence of dust covering factor on 
AGN luminosity.

A significant fraction (259/785, or 33\%) 
of our AGN candidates contained no evidence for an 
AGN in terms of emission line diagnostics, either because
the emission lines were detected, but did not have AGN-like
line ratios, or because the spectra had at most only a 
single weak emission line. 25 of these have 
some other evidence for AGN (X-ray emission or AGN-related radio emission),
an X-ray detection rate not much lower than the type-2 
quasars. Overall, the composite spectrum 
shows that the stellar population of their host galaxies seems similar
to those of the type-2 quasars. Most of them may therefore be 
be AGN whose narrow-line regions are obscured, but as we cannot prove this 
(except in $\approx 10$\% of cases where we have other diagnostics) they are
excluded from our AGN statistics.

This study reinforces the need for multiple AGN diagnostics to be employed
to complete a census of AGN, even for surveys of luminous AGN and quasars
such as this one. Nevertheless, our sample will provide important insights
into the luminosity dependence and cosmic evolution of obscuration. 
The publication of the WISE (Wright et al.\ 2010) all-sky survey 
will also allow us to continue
this survey into even brighter regimes, which will be particularly useful 
in understanding the luminosity dependence of obscuration at $z<1$.

\acknowledgments

NS is the receipient of an Australian Research Council Future Fellowship. 
ML, AP and TU were visiting astronomers 
at the Infrared Telescope Facility, which is 
operated by the University of Hawaii under cooperative agreement no. 
NNX-08AE38A with the National Aeronautics and Space Administration (NASA), 
Science Mission Directorate, Planetary Astronomy Program. 
The Gemini Observatory is operated by the Association of Universities for
Research in Astronomy, Inc, under a cooperative agreement with the
National Science Foundation on behalf of the Gemini Partnership.
Partly based on observations obtained at the Southern Astrophysical Research 
(SOAR) telescope, which is a joint project of the Minist\'{e}rio da 
Ci\^{e}ncia, Tecnologia, e Inova\c{c}\~{a}o (MCTI) da Rep\'{u}blica 
Federativa do Brasil, the U.S. National Optical Astronomy Observatory 
(NOAO), the University of North Carolina at Chapel Hill (UNC), and 
Michigan State University (MSU). 
We thank the Telescope System Instrumentation Program (TSIP), 
administered by NOAO, for the
opportunity to observe on the Multiple Mirror Telescope (MMT).
This research has made use of the 
NASA/IPAC Extragalactic Database (NED), which is operated by the 
Jet Propulsion Laboratory, California Institute of Technology, 
under contract with the National Aeronautics and Space Administration.
This research has made use of data obtained from the $Chandra$ Source Catalog,
provided by the $Chandra$ X-ray center (CXC) as part of the $Chandra$ Science
Archive. 
The NRAO is a facility of the National Science Foundation operated under 
cooperative agreement by Associated Universities, Inc.
This paper is dedicated to the late Steve Rawlings, whose work on radio
galaxy samples directly inspired the approach taken to this project.

\appendix

\section{Notes on individual objects}

\subsection{SW021749.00-052306.9}
This object had a redshift of 0.987 assigned by L07 on the basis of the
detection of a single emission line, presumed to be [OII]3727. Lanzuisi et al.\
(2009) obtained an improved spectrum of this object, showing that the redshift
is in fact 0.914 (consistent with the line we identified in L07 being 
[NeIII]3869 rather than [OII]. Lanzuisi et al.\ also detect strong [OIII]5007 
emission beyond the wavelength range of our original spectrum, confirming it 
as a type-2 AGN.

\subsection{SW021928.77-045433.7}
There is a clear detection of an emission line in the $H$-band spectrum 
from Gemini, but other features are less certain. The [NeV]3426
line is in a region of poor atmospheric transmission, so may be a
noise spike.
The optical spectrum, however, 
shows marginal evidence of a spectral break around
the wavelength of Ly$\alpha$ if the redshift is 3.23, and a marginally
significnant feature corresponding CIV1549 if the
line in $H$-band being [OII]3727. We therefore assign this object a 
redshift of 3.23 with a redshift quality of two.

\subsection{SW021822.13-050614.1}

This low redshift ($z=0.044$) object is an X-ray source. 
Optically it is classed as a non-AGN, 
but, as discussed by Severgnini et al.\ (2003),  its true nature is an AGN.

\subsection{SW021947.53-051008.5}

This quasar shows a blueshifted HeI line in absorption (Figure 3). 
Leighly, Dietrich \& Barber (2011) discuss a broad absorption line quasar with 
HeI absorption that may be similar to our object.

\subsection{SW022003.58-045145.6}
This object is a highly dust-reddened quasar at $z=1.33$ (Figure 3). 
The Hale and Gemini near-infrared spectra show only a single broad 
emission line at 1.53$\mu$m, however, a Keck/LRIS spectrum shows several 
UV emission lines at a redshift of $z=1.33$.

\subsection{SW104159.83+585856.4}
This otherwise unremarkable $z=0.342$ type-2 quasar has an 
anomalous emission line at 4331\AA$\;$which may correspond to an emission
line from a higher redshift galaxy lensed by the quasar. 

\subsection{SW105201.92+574051.5}
This object was independently identified by Coppin et al.\ (2010) as a 
high redshift, submm bright obscured 
quasar (AzLOCK.01 in that paper). 
The {\em Spitzer} IRS spectrum obtained by Coppin et al.
shows both strong PAH features and a strong continuum at $z\approx 2.5$. 
Our Gemini/NIRI spectrum (Figure 4) shows a narrow H$\alpha$ emission line at $z=2.467$.

\subsection{XFLS 171053.51+594433.1}
This object, the second-highest redshift quasar in the sample, was
first identified by Marleau et al.\ (2007) in a survey of radio
emitters in the XFLS. The quasar has a flux density of 2.4mJy at 
1.4GHz, so is radio-intermediate in nature. The optical spectrum
in figure 3 of Marleau et al.\ shows only narrow Ly$\alpha$ and NV1240, 
but our Hale spectrum with Triplespec (Figure 6) clearly shows a broad
H$\beta$ emission line, hence its classification as a reddened type-1 quasar.

\subsection{XFLS171419.9+602724}
This object, a $z=2.99$ type-2 quasar is discussed in detail in Lacy et al.\ 
(2011).

\subsection{XFLS171754.6+600913¯}
At $z=4.27$, this is the highest redshift object in our sample, and the most
luminous. It is a dust-reddened type-1 quasar with a mini-LoBAL system 
(Figure 6) with an overall luminosity of $\approx 10^{14}L_{\odot}$, and 
is a $Chandra$ X-ray detection, though falling well below the X-ray -- 15$\mu$m luminosity correlation, presumably due to absorption.

\end{document}